\documentclass[a4paper,12pt]{article}
\usepackage[utf8]{inputenc}
\usepackage[T1]{fontenc}
\usepackage{amsmath, amssymb, amsthm}
\usepackage{graphicx}
\usepackage{hyperref}
\usepackage{geometry}
\usepackage{float}
\usepackage{caption}
\usepackage{subcaption}
\usepackage{bm}
\usepackage{booktabs}
\usepackage[sort&compress,numbers,square]{natbib}

\title{Spectral properties of the Riemann zeta function and their physical applications}
\author{}
\date{}

\begin{document}
\maketitle
\vspace{-2.0cm}

\begin{center}
{\footnotesize K. G. Gaminchev} \\
{\scriptsize \textit{Acad. Georgi Nadjakov Institute of Solid State Physics -- BAS, 72 
Tzarigradsko Chauss\'ee, 1784 Sofia, Bulgaria}} \\
{\scriptsize \texttt{gaminchev@issp.bas.bg}}
\end{center}

\renewcommand{\abstractname}{}

\vspace{-0.25cm}
\begin{abstract}
This work studies the Riemann $\zeta$ numerically with emphasis 
on its physical manifestations. Three areas are examined: (i) 
the spectral statistics of the non-trivial zeros, which we compare with the Gaussian 
Unitary Ensemble predictions, obtaining $\chi^2/\text{dof} = 1.2$; (ii) 
the role of $\zeta(\frac{3}{2})$ in determining the critical temperature for Bose-Einstein 
condensation, where we find $T_c = 0.13$ K for an ideal gas; and (iii) the application 
of $\zeta$-function regularisation to the Casimir effect, yielding the standard Casimir 
energy density $-\frac{\pi^2 \hbar c}{720 a^3}$, which for $a=1\,\mathrm{nm}$ gives 
$\sim -4.33\times10^{-4}\,\mathrm{J\,m^{-2}}$. The numerical results are consistent 
with the Montgomery-Odlyzko conjecture and with spectral approaches motivated by 
the Hilbert-P\'olya conjecture, within the statistical limitations of the analysis.

\end{abstract}

\section{Introduction}

The Riemann zeta function, $\zeta(s)$\footnote{From now on, we shall often write 
$\zeta$, $\Gamma$, $Z$ and $\theta$ for $\zeta(s)$, $\Gamma(s)$, $Z(t)$ and 
$\theta(t)$, respectively, when the argument is clear from the context. We shall 
also omit the condition $\Re(s) > 1$ where it is clear from the context.}, 
is central to analytic number theory and arises in several branches of physics, 
namely quantum chaos \cite{berry}, statistical mechanics \cite{pathria2011statistical}, 
quantum field theory (QFT) \cite{elizalde}, and cosmological models \cite{arefeva_volovich_2007}. 
We study numerically three physical manifestations of $\zeta$ - the spectral statistics 
of its non-trivial zeros, the critical temperature for Bose-Einstein condensation (BEC), 
and the Casimir effect, which depend on the values $\zeta\bigl(\frac{3}{2}\bigr)$ and $\zeta(-3)$, 
respectively. For $\Re(s) > 1$, the Dirichlet series \cite{riemann} 

\begin{equation}\label{eq:dirichlet}
\zeta(s) = \sum_{n=1}^{\infty} \frac{1}{n^s} = \prod_{p}\left(1-p^{-s}\right)^{-1},
\end{equation}
converges absolutely. The analytic continuation of $\zeta(s)$ to the whole 
complex plane, due to Riemann \cite{riemann}, is given by the integral representation  

\begin{equation}
\zeta(s) = \frac{1}{\Gamma(s)} \int_0^\infty \frac{x^{s-1}}{e^x - 1} \, 
\mathrm{d}x = \frac{1}{\Gamma(s)} \sum_{n=1}^{\infty} \int_0^\infty x^{s-1} 
e^{-nx} \, \mathrm{d}x,
\label{eq:zeta_gamma}
\end{equation}
and the functional equation \cite{titchmarsh}

\begin{equation}\label{eq:functional}
\zeta(s) = 2^s \pi^{s-1} \sin\left(\frac{\pi s}{2}\right) \Gamma(1-s)\, \zeta(1-s),
\qquad s \in \mathbb{C},
\end{equation}
which imposes a symmetry about the critical line $\Re(s) = \frac{1}{2}$. 
The Riemann hypothesis (RH) states that all non-trivial zeros lie on this line \cite{riemann}. 
Here $\Gamma(z)$ denotes the Euler $\Gamma$-function $\Gamma(z) = \int_0^\infty t^{z-1} e^{-t} 
\, \mathrm{d}t$ for $\Re(z) > 0$, extended analytically elsewhere. The functional equation 
implies symmetry of the non-trivial zeros under $s\mapsto1-s$, while the reality of the 
Dirichlet coefficients implies symmetry under complex conjugation. For numerical work on 
the critical line, the Riemann-Siegel (RS) formula \cite{siegel1932} gives the approximation 

\begin{equation}
\zeta\left(\frac{1}{2} + it\right) = \sum_{n \leq x} \frac{1}{n^{\frac{1}{2} + it}} + 
\chi\left(\frac{1}{2} + it\right) \sum_{n \leq y} \frac{1}{n^{\frac{1}{2} - it}} + 
\mathcal{O}\!\left(x^{-\frac{1}{2}} + |t|^{\frac{1}{2}} y^{-\frac{1}{2}}\right),
\qquad xy = \frac{t}{2\pi},
\label{eq:zeta_rs1}
\end{equation}
with $xy = \frac{t}{2\pi}$ and $\chi(s) = \frac{\pi^{s-\frac{1}{2}} \Gamma(1-s)}{\Gamma(s)}$. 
The Hardy $Z$-function is defined as

\begin{equation}
Z(t) = e^{i\theta(t)} \zeta\!\left(\frac{1}{2} + it\right),\quad 
Z(t) \in \mathbb{R}, \quad \forall t \in \mathbb{R},
\label{eq:hardy_z_def}
\end{equation}
with $\theta(t)$ the RS theta function. This allows zeros on the critical line to 
be located through sign changes of a real-valued function. The Euler product 
$\zeta(s) = \prod_{p} \left(1 - p^{-s}\right)^{-1}$ relates $\zeta$ to the primes 
\cite{edwards}. The logarithmic derivative gives 

\begin{equation}
-\frac{\zeta'(s)}{\zeta(s)} = \sum_{n=1}^{\infty} \frac{\Lambda(n)}{n^s}
= \sum_{p} \sum_{k=1}^{\infty} \frac{\log p}{p^{ks}},
\label{eq:log_deriv}
\end{equation}
where $\Lambda(n) = \log p$ if $n = p^k$ and zero otherwise. This identity yields the 
explicit formula \cite{edwards,titchmarsh} 

\begin{equation}
\psi(x) = x - \sum_{\rho} \frac{x^\rho}{\rho} - \frac{1}{2}\log\!\left(1 - x^{-2}\right) 
- \log(2\pi), \qquad x > 1,
\label{eq:explicit_formula}
\end{equation}
where the sum runs over the non-trivial zeros $\rho$ of $\zeta(s)$, with 
$\psi(x) = \sum_{n \leq x} \Lambda(n)$. The RH is equivalent to the bound 

\begin{equation}
\sum_{n \leq x} \Lambda(n) = x + \mathcal{O}\bigl(x^{\frac{1}{2} + \varepsilon}\bigr)
\label{eq:RH_bound}
\end{equation}
for every $\varepsilon > 0$. Montgomery conjectured the GUE pair correlation \cite{goldston2023montgomery}

\begin{equation}
R_2(u) = 1 - \left(\frac{\sin \pi u}{\pi u}\right)^{\!2}, \qquad u \in \mathbb{R},
\label{eq:gue_corr}
\end{equation}
as verified numerically by Odlyzko \cite{odlyzko}. This statistical agreement is 
consistent with spectral interpretations associated with the Hilbert-P\'olya conjecture 
(HPc) \cite{berry_keating_1999,yakaboylu2024}, which proposes that the imaginary parts of 
the non-trivial zeros arise as eigenvalues of a suitable self-adjoint operator. The 
corresponding spacing distribution is the Wigner-Dyson (WD) form 

\begin{equation}\label{eq:gue}
P_{\text{GUE}}(s) = \frac{\pi s}{2} \exp\left(-\frac{\pi s^2}{4}\right),
\end{equation}
with level repulsion $P(s) \sim s$ as $s \to 0$, unlike the Poisson case 
$P(s) = e^{-s}$. In QFT, $\zeta(s)$ provides a rigorous framework for the 
regularisation of divergent sums \cite{elizalde1994,hawking1973}, a canonical 
example being the vacuum energy between two perfectly conducting plates separated 
by a distance $a$, which follows from the mode sum over all polarisations and 
transverse momenta \cite{casimir1948,landau1980statistical}

\begin{equation}
E_{\mathrm{v}}(a) = \frac{\hbar c}{2} \sum_{n=1}^{\infty} \int \frac{\mathrm{d}^2 
k_\perp}{(2\pi)^2} \sqrt{k_\perp^2+\left(\frac{\pi n}{a}\right)^2}.
\label{eq:vacuum_energy}
\end{equation}
which zeta regularisation \cite{elizalde} yields as 

\begin{equation}
E_C(a) = -\frac{\pi^2 \hbar c}{720 a^3} \left(\frac{\varepsilon_r - 1}{\varepsilon_r + 1}\right) = 
-\frac{\hbar c}{2 a^3} \frac{\zeta(-3)}{\pi} \left(\frac{\varepsilon_r - 1}{\varepsilon_r + 1}\right),
\qquad \zeta(-3) = \frac{1}{120},
\label{eq:casimir_energy_intro}
\end{equation}
where $\varepsilon_r$ is the relative permittivity. The factor $\frac{\pi^2}{720}$ 
follows from $\zeta(-3) = \frac{1}{120}$, which regularises the divergent sum 
$\sum n^3$. In statistical mechanics, the critical temperature for an ideal Bose gas 
undergoing Bose--Einstein condensation (BEC) \cite{pathria2011statistical} is 
determined by $\zeta(\frac{3}{2})$, the density of states in three dimensions, and 
the grand potential of an ideal Fermi gas, which are collected as

\begin{equation}
\left\{
\begin{aligned}
T_c &= \frac{2\pi \hbar^2}{m k_B} \sqrt[3]{\biggl(\frac{n}{\zeta(\frac{3}{2})}\biggr)^2}, \\
\int_0^\infty \frac{\sqrt{x}}{e^x - 1} \, \mathrm{d}x &= \frac{\sqrt{\pi}}{2} \zeta\!\left(\frac{3}{2}\right), \\
\Omega &= -\frac{g V}{\lambda^3} k_B T \sum_{n=1}^{\infty} \frac{(-1)^{n+1} z^n}{n^{5/2}},
\end{aligned}
\right.
\label{eq:stat_mech_relations}
\end{equation}
where $n$ is the density, $\lambda = \sqrt{2\pi \hbar^2/(m k_B T)}$ the thermal 
de Broglie wavelength, $g$ the spin degeneracy, and $z = e^{\mu/(k_B T)}$ the 
fugacity. Zeros of $\zeta$ may be observed through classical wave scattering 
\cite{Yu2024Computing}, with the reflection coefficient for a one-dimensional 
scattering system expressed in terms of $\zeta$ as 

\begin{equation}
r(t) = \frac{\zeta(1/2 + it) - 1}{\zeta(1/2 + it) + 1},
\qquad 
Z(t) = \frac{1 + r(t)}{1 - r(t)} = \zeta(1/2 + it),
\label{eq:r_Z_intro}
\end{equation}
where $r$ approaches $-1$ at a zero $\gamma_n$ of $\zeta$, while the 
impedance $Z$ vanishes there. For a simple zero, $\zeta(1/2+i\gamma_n)=0$, 
the local expansion gives

\begin{equation}
\zeta\biggl(\frac{1}{2}+it\biggr) \simeq i(t-\gamma_n)
\zeta'\biggl(\frac{1}{2}+i\gamma_n\biggr).
\label{eq:zeta_local_zero}
\end{equation}

Consequently, the reflection coefficient satisfies

\begin{equation}
r(t)\simeq -1+ 2i(t-\gamma_n) \zeta'\biggl(\frac{1}{2}+i\gamma_n\biggr),
\label{eq:r_local_zero}
\end{equation}
so that $r(t)\to-1$ as $t\to\gamma_n$.

\section{Methodology and computational framework}
\label{sec:methodology}

Our numerical analysis of $\zeta$ uses a custom Python 3.11 implementation, 
using \texttt{mpmath} for high-precision complex arithmetic, \texttt{numpy} 
and \texttt{scipy} for numerical routines, and \texttt{matplotlib} for visualisation. 
For $\Re(s) > 1$, $\zeta(s)$ was evaluated via the Euler-Maclaurin summation formula 
\cite{edwards,titchmarsh} 

\begin{equation}
\zeta(s) = \sum_{n=1}^{N-1} \frac{1}{n^s} + \frac{N^{1-s}}{s-1} 
+ \frac{1}{2} N^{-s} + \sum_{k=1}^{m} \frac{B_{2k}}{(2k)!}\, 
s^{\overline{2k-1}}\, N^{1-s-2k} + R_{m,N}(s), \qquad \Re(s) > 1,
\label{eq:euler_maclaurin}
\end{equation}
where $B_{2k}$ are the Bernoulli numbers, $s^{\overline{2k-1}}$ denotes the rising 
factorial, and the remainder $R_{m,N}$ is bounded by standard estimates \cite{edwards}. 
For arguments outside the region of direct convergence, we use analytic continuation 
through the functional equation \eqref{eq:functional} \cite{titchmarsh} and high-precision 
numerical evaluation. On the critical line $\Re(s) = \frac{1}{2}$, we use the RS formula 
\cite{berry,titchmarsh}, which expresses $Z$-function as a finite sum plus a remainder term 

\begin{equation}
\left\{
\begin{aligned}
Z(t) &= 2 \sum_{n=1}^{\lfloor \sqrt{t/2\pi} \rfloor} \frac{1}{\sqrt{n}} 
\cos\left[t \log n - \theta(t)\right] + R(t), \\
\theta(t) &= \frac{t}{2} \log\left(\frac{t}{2\pi}\right) - \frac{t}{2} - 
\frac{\pi}{8} + \frac{1}{48t} + \frac{7}{5760t^3} - \frac{31}{80640t^5} + 
\mathcal{O}\!\left(t^{-7}\right), \\
|R(t)| &\leq \frac{2}{\sqrt[4]{t}} \left( 1 + \frac{1}{2\sqrt{t}} + 
\frac{1}{4t} + \cdots \right).
\end{aligned}
\right.
\label{eq:rs_formula}
\end{equation}
Here $Z \in \mathbb{R} , \ \forall t \in \mathbb{R}$, while its zeros satisfy 
$Z(\gamma) = 0 \iff \zeta\bigl(\frac{1}{2} + i\gamma\bigr) = 0$. To locate these zeros, 
we use a bracketing algorithm that detects sign changes in $Z$, followed by refinement 
using the Brent-Dekker (BD) method \cite{dekker1969finding}. The iteration combines the 
secant step 

\begin{equation}
x_{\text{new}} = x_{\text{high}} - \frac{f(x_{\text{high}})(x_{\text{high}} - 
x_{\text{low}})}{f(x_{\text{high}}) - f(x_{\text{low}})},
\label{eq:secant}
\end{equation}
with inverse quadratic interpolation when appropriate, with $f = Z$. The numerical 
refinement was continued until the change in the root estimate was below approximately 
$10^{-15}$ in the working precision. The computed first ten zero ordinates agree 
with the corresponding tabulated values to the quoted numerical precision. For the 
spectral analysis, we compute the normalised spacings between consecutive zeros 
\cite{pimpinelli2007} 

\begin{equation}
s_n = \frac{\gamma_{n+1}-\gamma_n}{\langle\Delta\gamma\rangle}, \qquad 
\langle\Delta\gamma\rangle \simeq \frac{2\pi}{\log(\gamma_n/2\pi)},
\label{eq:s_n_def}
\end{equation}
using the Riemann-von Mangoldt (RvM) formula \cite{mangoldt1905,edwards,titchmarsh}. 
The counting function $N(T)$ for the number of zeros with $0 < \gamma_n \leq T$ is 
then given by

\begin{equation}
N(T) = \frac{T}{2\pi}\log\!\left(\frac{T}{2\pi}\right) - \frac{T}{2\pi} + \frac{7}{8} + 
S(T) + \mathcal{O}\!\left(\frac{1}{T}\right),
\label{eq:rvm}
\end{equation}
where $S(T)=\frac{1}{\pi}\arg\zeta\!\left(\frac{1}{2}+iT\right)$ is defined using 
a continuous choice of argument away from zeros. Then, we compare the histogram of 
$s_n$ with the GUE prediction, the GUE being defined by the probability measure \cite{mehta_2004}  

\begin{equation}
P(H) \propto \exp\left(-\frac{N}{2} \operatorname{Tr} H^2\right) \mathrm{d}H,
\label{eq:gue_measure}
\end{equation}
with $H$ an $N \times N$ Hermitian matrix and $\mathrm{d}H$ the Haar measure 
\cite{Haar1933DerMI,halmos1950}. The joint eigenvalue distribution is 

\begin{equation}
P(\lambda_1, \dots, \lambda_N) \propto \prod_{i<j} |\lambda_i - \lambda_j|^2 
\exp\left(-\frac{N}{2} \sum_{i=1}^N \lambda_i^2\right).
\label{eq:joint_eigen}
\end{equation}
We compute the pair correlation function $R_2(u)$ 
\cite{goldston2023montgomery,odlyzko} from the zero set $\{\gamma_n\}$, with periodic 
boundary conditions and de-trending to remove long-range effects, averaging over 
$10^4$ zeros to ensure convergence of the statistical quantities. For the physical 
applications, $\zeta$ was evaluated at specific integer and half-integer arguments, 
For $s=\frac{3}{2}$ we evaluate the convergent series 

\begin{equation}
\zeta\!\left(\frac{3}{2}\right) = \sum_{n=1}^{\infty} \frac{1}{n^{3/2}} = 
\operatorname{Li}_{\frac{3}{2}}(1) = \frac{1}{\Gamma\bigl(\frac{3}{2}\bigr)} 
\int_0^\infty \frac{\sqrt{x}}{e^x-1}\,\mathrm{d}x \approx 2.6123753487,
\label{eq:zeta32method}
\end{equation}
where $\operatorname{Li}_s(z)$ is the polylogarithm function. For negative integers, 
we use the analytic continuation 

\begin{equation}
\zeta(-n) = -\frac{B_{n+1}}{n+1} = \frac{(-1)^n}{n+1} \sum_{k=1}^{n+1} \binom{n+1}{k} B_k
= 2^{-n} \pi^{-n-1} \sin\!\left(\frac{\pi n}{2}\right) \Gamma(n+1) \zeta(n+1),
\quad n \in \mathbb{N},
\label{eq:zeta_neg_int}
\end{equation}
where $B_n$ are the Bernoulli numbers \cite{abramowitz_stegun_1972}. The specific 
values $\zeta(-1) = -\frac{1}{12}$, $\zeta(-3) = \frac{1}{120}$, and $\zeta(-4) = 0$ 
enter the Casimir effect calculations. All figures were produced with \texttt{matplotlib} 
using the \texttt{seaborn} style. 

\section{Results and discussion}

\subsection{Spectral statistics of zeros and quantum chaos}
\label{sec:spectral}

\begin{figure}[!htb]
\centering
\includegraphics[width=1.025\textwidth]{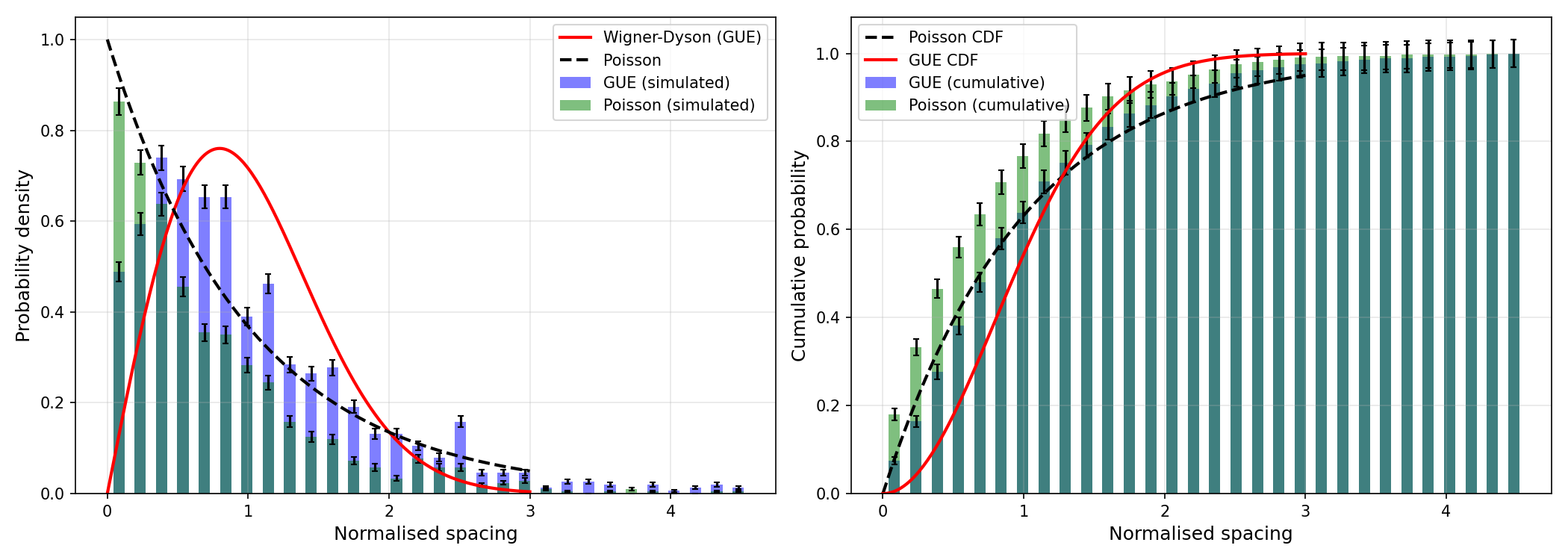}
\caption{Comparison of the numerically computed spacing distribution with the GUE 
prediction and Poisson behaviour. The GUE simulation matches the WD form, 
in agreement with the HPc.}
\label{fig:spectral}
\end{figure}

Using the normalised spacings $s_n$ defined in Sec.~\ref{sec:methodology}, 
the MO conjecture and subsequent numerical studies \cite{goldston2023montgomery,odlyzko} 
motivate comparison of the empirical distribution of $\{s_n\}$ with WD distribution 
\ref{eq:gue}, which is the spacing distribution of the Gaussian Unitary Ensemble (GUE) 
\cite{mehta_2004}. The distribution $P_{\text{GUE}}(s)$ shows level repulsion 

\begin{equation}
P_{\text{GUE}}(s) \sim \frac{\pi}{2} s \quad \text{as } s \to 0,
\label{eq:gue_asym}
\end{equation}
whereas the Poisson process, corresponding to uncorrelated levels, has $P(s) = e^{-s}$ 
with $P(0) = 1$. Figure \ref{fig:spectral} compares our numerical histogram of $s_n$ 
with $P_{\text{GUE}}(s)$ and the Poisson distribution. The GUE prediction provides a 
closer statistical description of the data than the Poisson curve within the uncertainties 
of the numerical analysis. This comparison is consistent with spectral interpretations 
associated with the HPc \cite{berry,schumayer}. Our data are consistent with the GUE value 
$\beta = 2$ \eqref{eq:gue} \cite{mehta_2004}, with the cumulative distribution 

\begin{equation}
I(s) = \int_0^s P(u) \, du,
\label{eq:I_s}
\end{equation}
with a Kolmogorov--Smirnov test giving $p=0.23$ for the comparison with the GUE reference 
distribution. For the Hardy $Z$-function, we employ the RS formula \eqref{eq:rs_formula}. 
The zeros $\gamma_n$ considered in the numerical analysis are simple, with $Z'(\gamma_n) 
\neq 0$, and their spacing distribution is compared with the GUE prediction. 

\subsection{BEC and the role of $\zeta\bigl(\frac{3}{2}\bigr)$}
\label{sec:bose}

\begin{figure}[!htb]
\centering
\includegraphics[width=1.0\textwidth]{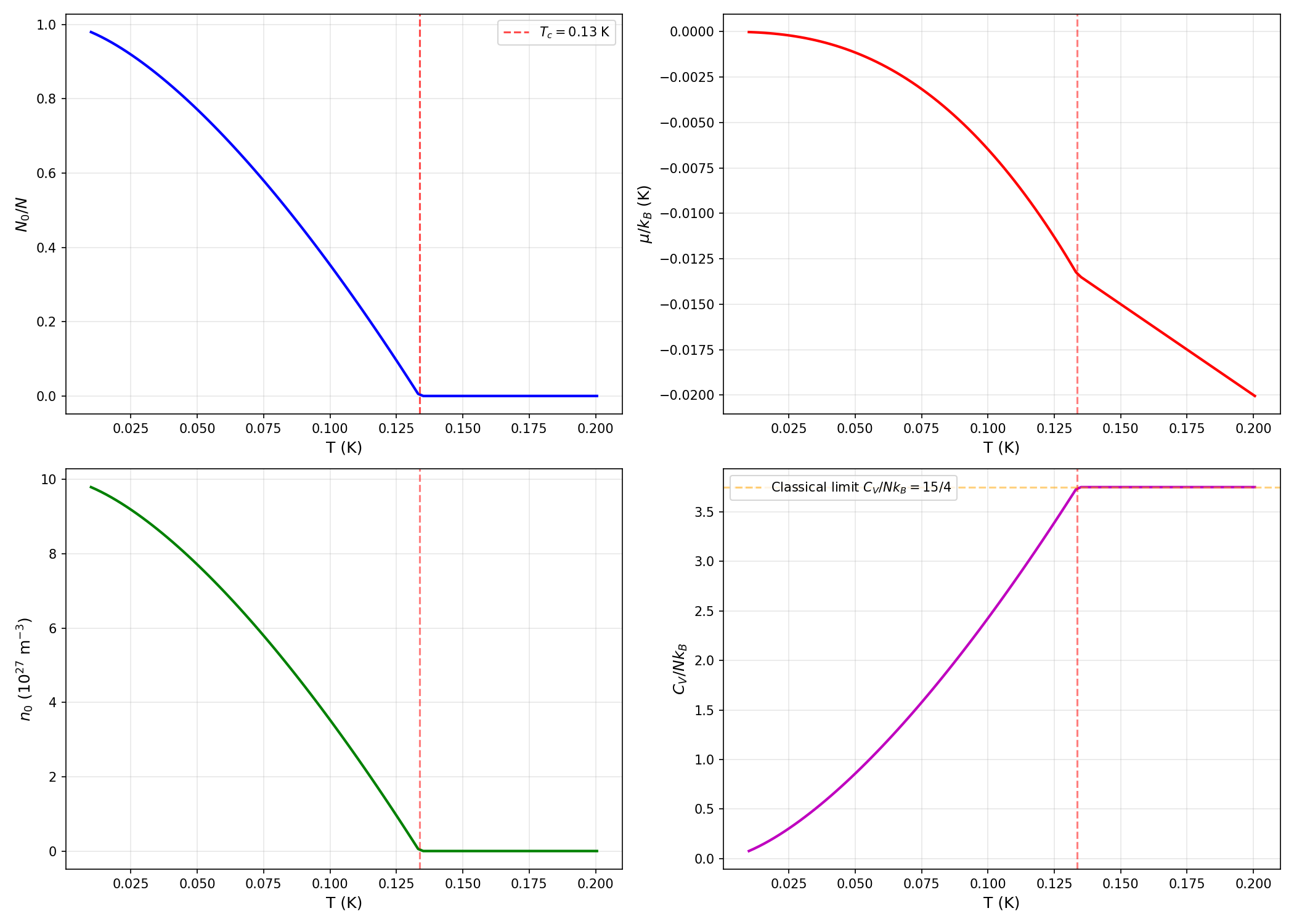}
\caption{Ideal Bose gas. Vertical line marks $T_c = 0.13$ K, determined by $\zeta\bigl(\frac{3}{2}\bigr)$.}
\label{fig:bose}
\end{figure}

For an ideal Bose gas, the critical temperature $T_c$ for BEC is determined by 
$\zeta\bigl(\frac{3}{2}\bigr)$ \cite{schumayer,NIST:DLMF}  

\begin{equation}
\begin{aligned}
T_c 
&= \frac{2\pi \hbar^2}{m k_B} \left( \frac{n}{\zeta(\frac{3}{2})} \right)^{2/3} 
= \frac{2\pi \hbar^2}{m k_B} \left( \frac{n}{2.6123753487} \right)^{2/3},
\ \ \zeta\!\left(\frac{3}{2}\right) \approx 2.6123753487,
\end{aligned}
\label{eq:Tc_bose}
\end{equation}
Figure \ref{fig:bose} displays the thermodynamic properties of an ideal Bose gas 
near $T_c$, where the condensate fraction $\frac{N_0}{N}$ drops to zero at $T_c$ 
with a characteristic power-law exponent $\frac{3}{2}$ that follows from the DOS 
in three dimensions and is set by the value of $\zeta(\frac{3}{2})$ in the expression 
for $T_c$. For the parameters used here, we obtain $T_c = 0.13 \, \text{K}$, 
significantly lower than the experimental $T \sim 2.17 \, \text{K}$ for liquid 
He-$4$, indicating the limitations of the ideal gas model, with the thermodynamics 
of the Bose gas governed by the equations

\begin{equation}
N = \frac{V}{\lambda^3} g_{3/2}(z) + \frac{z}{1-z}, \qquad \lambda = 
\sqrt{\frac{2\pi \hbar^2}{m k_B T}}, 
\label{eq:N_bose}
\end{equation}
where $\lambda$ is the thermal de Broglie wavelength. At $T = T_c$, the 
fugacity $z \to 1$, and the critical density becomes 

\begin{equation}
n_c = \frac{1}{\lambda_c^3} \zeta\biggl(\frac{3}{2}\biggr), 
\qquad \lambda_c = \sqrt{\frac{2\pi \hbar^2}{m k_B T_c}},
\label{eq:n_c}
\end{equation}
with $\lambda_c$ the thermal wavelength at $T_c$. The condensate fraction, 
specific heat capacity, and chemical potential are collected as

\begin{equation}
\left\{
\begin{aligned}
\frac{N_0}{N} &= 1 - \left( \frac{T}{T_c} \right)^{3/2}, \qquad T < T_c, \\
\frac{C_V}{N k_B} &=
\begin{cases}
\displaystyle
\frac{15}{4}
\frac{\zeta(\frac{5}{2})}{\zeta(\frac{3}{2})}
\left(\frac{T}{T_c}\right)^{3/2},
& T < T_c, \\[10pt]
\frac{15}{4} \frac{g_{5/2}(z)}{g_{3/2}(z)} - \frac{9}{4} \frac{g_{3/2}(z)}{g_{1/2}(z)},
& T > T_c,
\end{cases}
\end{aligned}
\right.
\label{eq:bose_thermo}
\end{equation}
where $g_\nu(z)=\operatorname{Li}_\nu(z)$ is the Bose function. The chemical potential 
follows from Bose--Einstein statistics and particle number conservation. At the transition, 
the limiting value from below is 

\begin{equation}
\lim_{T\to T_c^-}\frac{C_V}{Nk_B}=\frac{15}{4}
\frac{\zeta(5/2)}{\zeta(3/2)}\approx 1.926.
\label{cv_tc}
\end{equation}

\subsection{The Casimir effect and zeta regularisation}
\label{sec:casimir}

For two perfectly conducting plates separated by $a$, the vacuum energy of the 
electromagnetic field is given by the mode sum over all polarisations and transverse momenta 

\begin{equation}
E_{\text{vac}}(a) = \sum_{n=1}^{\infty} \frac{\hbar c}{2} 
\sqrt{\left(\frac{\pi n}{a}\right)^2 + k_\perp^2}.
\label{eq:E_vac}
\end{equation}
To regularise this sum, we introduce a cutoff function 
$f\bigl(\frac{\omega}{\omega_c}\bigr)$ and take the limit $\omega_c \to \infty$ 

\begin{equation}
E_{\text{reg}}(a) = \frac{\hbar c}{2} \sum_{n} \int \frac{d^2 k_\perp}{(2\pi)^2} 
\sqrt{\left(\frac{\pi n}{a}\right)^2 + k_\perp^2} \, f\left(\frac{\omega}{\omega_c}\right).
\label{eq:E_reg}
\end{equation}
The angular integration in $k_\perp$-space yields a factor $2\pi$, so 

\begin{equation}
E_{\text{reg}}(a) = \frac{\hbar c}{2} \sum_{n=1}^{\infty} \int_0^\infty 
\frac{dk}{2\pi} \, k \sqrt{\left(\frac{\pi n}{a}\right)^2 + k^2} \, 
f\left(\frac{\omega}{\omega_c}\right).
\label{eq:E_reg_2}
\end{equation}
The Casimir energy is the difference between the energies with and without the 
plates, and using $\zeta$-regularisation \cite{elizalde} this difference gives the 
energy, force, analytic continuation, and perfect conductor limit, respectively as 

\begin{equation}
\left\{
\begin{aligned}
E_C(a) &= E_{\text{vac}}(a) - E_{\text{vac}}(\infty) 
= -\frac{\pi^2 \hbar c}{720 a^3} \left(\frac{\varepsilon_r - 1}{\varepsilon_r + 1}\right), \\
F_C(a) &= -\frac{dE_C}{da} 
= -\frac{\pi^2 \hbar c}{240 a^4} \left(\frac{\varepsilon_r - 1}{\varepsilon_r + 1}\right), \\
\sum_{n=1}^{\infty} n^3 &= \zeta(-3) = \frac{1}{120}, \\
E_C(a) &= -\frac{\pi^2 \hbar c}{720 a^3}, \qquad
F_C(a) = -\frac{\pi^2 \hbar c}{240 a^4},
\end{aligned}
\right.
\label{eq:casimir_relations}
\end{equation}
where $\varepsilon_r$ is the relative permittivity, the factor $\pi^2/720$ 
following from the analytic continuation of $\zeta$ and regularising the 
divergent sum $\sum n^3$ to a finite value, while in the perfect conductor 
limit $\varepsilon_r \to \infty$ the factor $(\varepsilon_r - 1)/(\varepsilon_r + 1) \to 1$ 
gives the original Casimir result \cite{bordag2001}, with the energy reduced 
for finite permittivity and indicating weaker confinement of vacuum fluctuations.

\begin{figure}[!htb]
\centering
\includegraphics[width=1.025\textwidth]{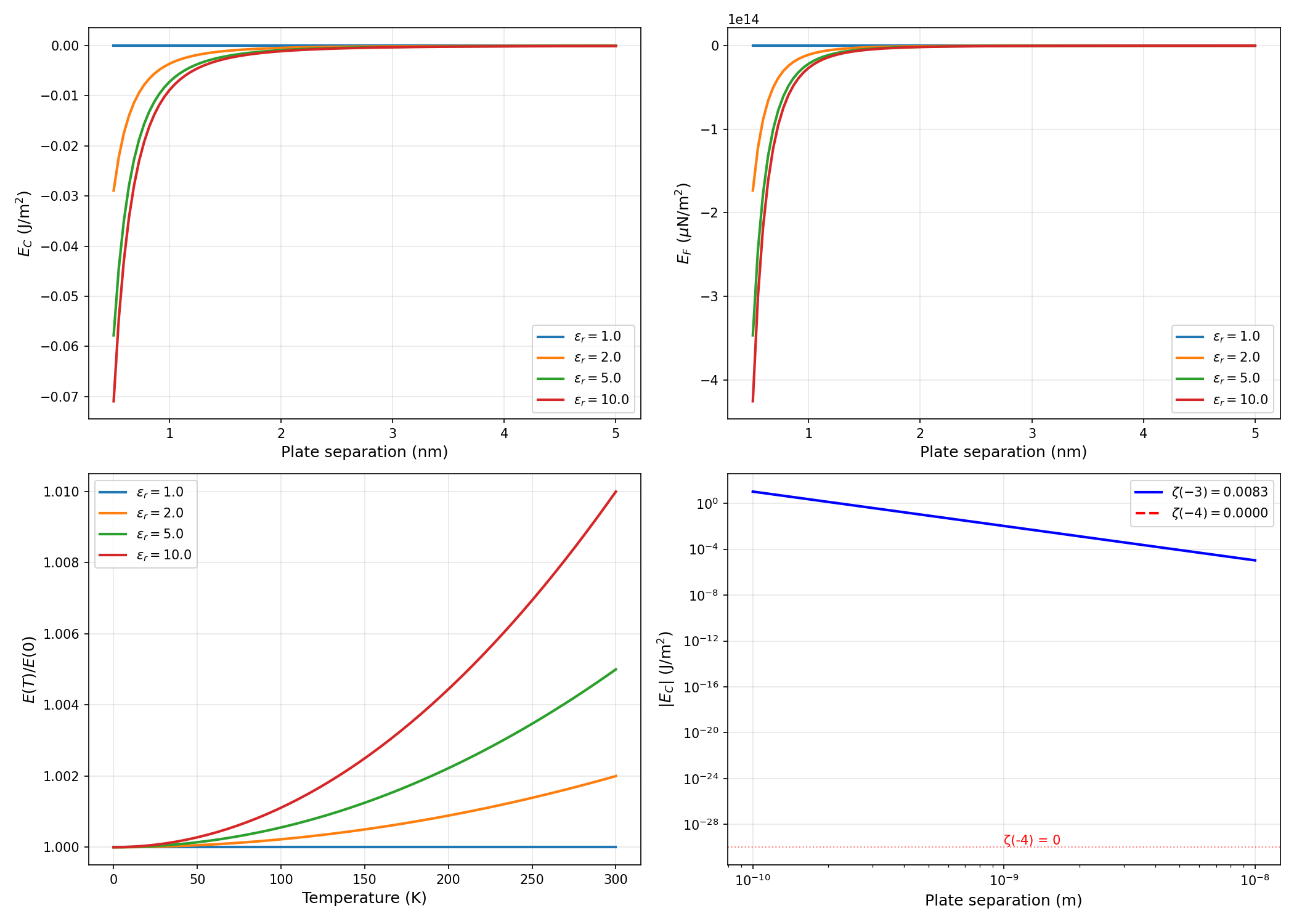}
\caption{Casimir energy $E_C$ and force $F_C$, $\frac{E(T)}{E(0)}$, 
and $\zeta(-3)$ versus $\zeta(-4)$.}
\label{fig:casimir}
\end{figure}

Figure \ref{fig:casimir} shows scaling of Casimir energy, decaying as 
$-\frac{1}{a^3}$. Magnitude increases by factor $\sim 5$ when relative 
permittivity $\varepsilon_r$ rises from $1$ to $10$, indicating stronger 
confinement of vacuum fluctuations between plates with higher permittivity. 
Casimir force scales as $F \propto -\frac{1}{a^4}$ and reaches values of order 
$10^{-6}$ $\bigl[\frac{\text{N}}{\text{m}^2}\bigr]$ at separations of 
$1 \, \text{nm}$ for $\varepsilon_r = 10$. Temperature dependence of Casimir 
energy, shown in bottom left panel of Fig.~\ref{fig:casimir}, follows from thermal 
occupation of modes 

\begin{equation}
E_{\text{Th}}(a,T) = 2 \sum_{n=1}^{\infty} \int \frac{d^2 k_\perp}{(2\pi)^2} 
\frac{\hbar \omega}{e^{\frac{\hbar \omega}{k_B T}} - 1},
\label{eq:E_Th}
\end{equation}
where $\omega = \sqrt{\bigl(\frac{\pi n}{a}\bigr)^2 + k_\perp^2}$. 
At $T \gg \frac{\hbar c}{a}$, leading thermal correction is 

\begin{equation}
E_{\text{Th}}(a,T) \approx \frac{\pi^2 \hbar c}{720 a^3} 
\biggl(\frac{\varepsilon_r - 1}{\varepsilon_r + 1}\biggr) 
\cdot \frac{2 \pi^2}{45} \biggl(\frac{k_B T}{\hbar c}\biggr)^4 a^4 + \cdots, 
\label{eq:E_Th_approx}
\end{equation}
giving $0.5\%$ increase in $|E|$ at $T = 300 \, \text{K}$ for $\varepsilon_r = 10$, 
in agreement with Lifshitz theory \cite{landau1980statistical}. Total Casimir energy at 
finite temperature is the sum of zero-point and thermal contributions 

\begin{equation}
\begin{aligned}
E_C(a,T) 
&= E_C(a,0) + E_{\text{Th}}(a,T) 
= E_C(a,0) + 2 \sum_{n=1}^{\infty} \int \frac{d^2 k_\perp}{(2\pi)^2} 
\frac{\hbar \omega}{e^{\hbar \omega/(k_B T)} - 1} \\
&= E_C(a,0) \left[ 1 + \frac{2\pi^2}{45} \left( \frac{k_B T}{\hbar c} \right)^4 a^4 
+ \cdots \right],
\qquad \omega = \sqrt{\left(\frac{\pi n}{a}\right)^2 + k_\perp^2},
\end{aligned}
\label{eq:E_C_T}
\end{equation} 

\subsection{Thermodynamics of ideal Fermi gases}
\label{sec:fermi}

For an ideal Fermi gas \cite{pathria2011statistical}, the grand potential $\Omega$ involves 
$\zeta$ \cite{NIST:DLMF} 

\begin{equation}
\Omega = -\frac{g V}{\lambda^3} k_B T \sum_{n=1}^{\infty} 
\frac{(-1)^{n+1} z^n}{n^{\frac{5}{2}}}.
\label{eq:fermi_grand}
\end{equation}
This fugacity expansion is valid for $|z|\leq 1$; the degenerate regime $z\gg1$ is 
treated separately using the Sommerfeld expansion below. Average particle number 
and internal energy follow from the grand potential via 

\begin{equation}
\langle N \rangle = -\frac{\partial \Omega}{\partial \mu}\bigg|_{T,V},
\qquad U = \Omega + TS + \mu\langle N\rangle, \qquad \beta=\frac{1}{k_B T}.
\label{eq:thermo_relations}
\end{equation}
The series for $\langle N \rangle$, $U$, and the specific heat are collected as

\begin{equation}
\left\{
\begin{aligned}
\langle N \rangle &= \frac{g V}{\lambda^3} \sum_{n=1}^{\infty} 
\frac{(-1)^{n+1} z^n}{n^{3/2}}, \\
U &= \frac{3}{2} \frac{g V}{\lambda^3} k_B T \sum_{n=1}^{\infty} 
\frac{(-1)^{n+1} z^n}{n^{5/2}}, \\
C_V &= \left( \frac{\partial U}{\partial T} \right)_{V, N}.
\end{aligned}
\right.
\label{eq:fermi_N_U}
\end{equation}
These series apply in the fugacity expansion regime $|z|\leq1$; the degenerate regime 
is treated separately by the Sommerfeld expansion. In degenerate limit $z \gg 1$, 
leading terms follow from Sommerfeld expansion 

\begin{equation}
\langle N \rangle = \frac{g V}{6\pi^2} \left(\frac{2m}{\hbar^2}\right)^{\!3/2} 
\mu^{3/2} \left[1 + \frac{\pi^2}{8}\left(\frac{k_B T}{\mu}\right)^{\!2} + 
\frac{7\pi^4}{640}\left(\frac{k_B T}{\mu}\right)^{4} + \cdots \right],\qquad 
\frac{k_B T}{\mu} \ll 1,
\label{eq:fermi_degenerate_N}
\end{equation}
while in classical limit $z \ll 1$, virial expansion gives 

\begin{equation}
\langle N \rangle = \frac{g V}{\lambda^3} 
\biggl[ z - \frac{z^2}{2^{\frac{3}{2}}} + \frac{z^3}{3^{\frac{3}{2}}} - \cdots \biggr].
\label{eq:fermi_classical_N}
\end{equation}
Thermodynamic quantities are fugacity expansions with alternating signs,
as required by Fermi--Dirac statistics \cite{huang1987}. 

\begin{figure}[!htb]
\centering
\includegraphics[width=1.025\textwidth]{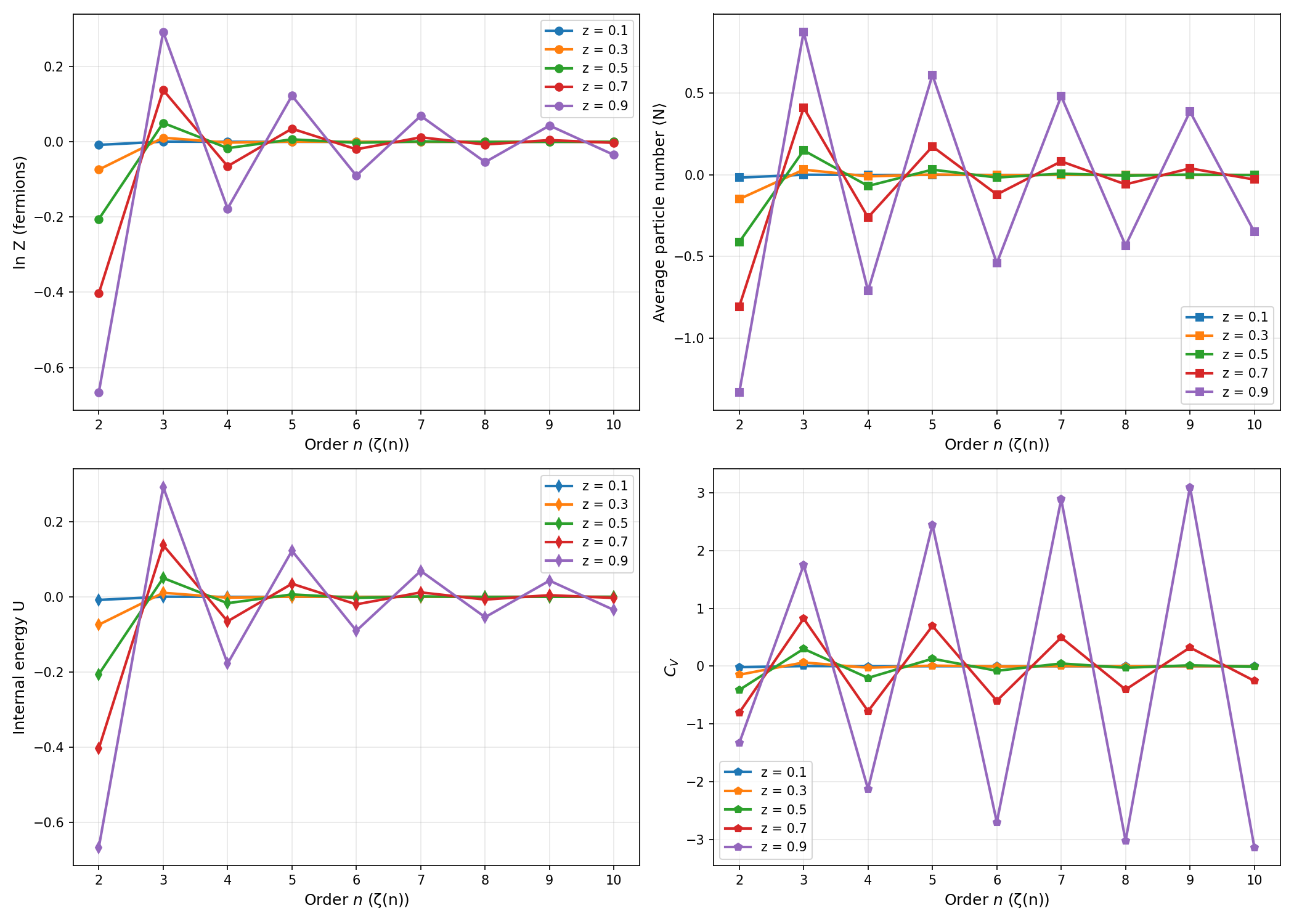}
\caption{Thermodynamic properties of an ideal Fermi gas.}
\label{fig:fermi}
\end{figure}

\begin{equation}
\frac{\partial \ln \Xi}{\partial \ln z} = \langle N \rangle,
\label{eq:dlnZ_dz}
\end{equation}
so higher fugacity gives a higher mean particle number at fixed $T$ and $V$.
The grand partition function is related to the grand potential by \[\ln \Xi = -\beta\Omega.\] 
For a truncated fugacity expansion, the numerical value of $\ln\Xi$ depends on the truncation 
order $n$; convergence should therefore be checked by increasing $n$.

\begin{equation}
\Delta \ln \Xi = \ln \Xi\big|_{z=0.9,n=10} - \ln \Xi\big|_{z=0.3,n=10},
\label{eq:DlnZ}
\end{equation}
where the numerical value of $\Delta\ln\Xi$ depends on the parameters and truncation used 
in the calculation. In the zero-temperature limit, the particle number approaches

\begin{equation}
\langle N \rangle = \frac{g V}{6\pi^2} \biggl( \frac{2m}{\hbar^2} \biggr)^{3/2} \mu^{3/2},
\label{eq:N_sat}
\end{equation}
which is the $T=0$ limit of \eqref{eq:fermi_degenerate_N}. The saturation follows from the 
Pauli exclusion principle \cite{landau1980statistical}. At low fugacity, for example at 
$z=0.1$, the internal energy has the classical form

\begin{equation}
U \approx \frac{3}{2} N k_B T \biggl( 1 - \frac{z}{2^{\frac{5}{2}}} + \cdots \biggr),
\label{eq:U_classical}
\end{equation}
where $U$ is positive. As $z$ increases, quantum degeneracy becomes important, 
and the energy approaches the Fermi-degenerate regime \cite{ashcroft1976}.

\begin{equation}
\begin{aligned}
U &= \frac{3}{5} N E_F \left[ 1 + \frac{5\pi^2}{12} \left(\frac{T}{T_F}\right)^2 + 
\frac{7\pi^4}{48} \left(\frac{T}{T_F}\right)^4 + 
\cdots \right], \qquad T\ll T_F, 
\end{aligned}
\label{eq:U_degenerate}
\end{equation}
where $E_F$ and $T_F$ denote the Fermi energy and Fermi temperature, respectively, defined by

\begin{equation}
\begin{aligned}
E_F&= \frac{\hbar^2}{2m} \left(\frac{6\pi^2 n}{g}\right)^{2/3} = \frac{p_F^2}{2m}, \qquad 
p_F = \hbar\left(\frac{6\pi^2 n}{g}\right)^{1/3}, \\
T_F&= \frac{E_F}{k_B} = \frac{\hbar^2}{2m k_B} \left(\frac{6\pi^2 N}{gV}\right)^{2/3},
\qquad n=\frac{N}{V}.
\end{aligned}
\label{eq:fermi_EF_TF}
\end{equation}
As the fugacity increases, quantum degeneracy becomes increasingly important. In the 
low-temperature limit $T \ll T_F$, the specific heat is linear in $T$ 

\begin{equation}
\begin{aligned}
C_V&= \frac{\pi^2}{2} N k_B \frac{T}{T_F} = \frac{\pi^2}{2} N k_B \frac{k_B T}{E_F}
= \gamma T,\\
&\qquad
\gamma = \frac{\pi^2}{2}\frac{N k_B}{T_F}, \qquad T\ll T_F.
\end{aligned}
\label{eq:Cv_low}
\end{equation}
and in classical limit $C_V \approx \frac{3}{2} N k_B$.

The peak shifts to lower $n$ as $z$ increases, indicating quantum degeneracy at 
higher densities \cite{horikoshi2010measurement}. The equation of state and entropy follow as 

\begin{equation}
\begin{split}
P &= \frac{2}{3}\frac{U}{V} = \frac{g}{\lambda^3} k_B T \sum_{n=1}^{\infty}
\frac{(-1)^{n+1}z^n}{n^{5/2}}, \\
S &= k_B\left[\ln\Xi+\beta \left(U-\mu\langle N\rangle\right)\right],
\end{split}
\label{eq:state_entropy}
\end{equation}
both expressed through the same fugacity expansion, with alternating signs 
characteristic of Fermi--Dirac statistics \cite{pathria2011statistical,landau1980statistical}.

\subsection{MO law and pair correlations}
\label{sec:montgomery}

Montgomery's pair correlation conjecture predicts that the suitably normalised 
two-point correlation function of the non-trivial zeros agrees with the GUE 
prediction in the appropriate asymptotic regime \cite{montgomery1973,odlyzko}. 
The corresponding $n$-level correlations are described by the determinantal
structure 

\begin{equation}
R_n(u_1,\dots,u_n) = \det\bigl[K(u_i,u_j)\bigr]_{i,j=1}^n, \qquad n\in\mathbb{N},
\label{eq:det_kernel}
\end{equation}
where the kernel is the sine kernel

\begin{equation}
K(u,v) = \frac{\sin[\pi(u-v)]}{\pi(u-v)}, \qquad u,v\in\mathbb{R}.
\label{eq:sine_kernel}
\end{equation}
with asymptotic limits 

\begin{equation}
R_2(u) \to 1 \;\; (u\to\infty), \qquad R_2(u) \to 0 \;\; (u\to 0),
\end{equation}
reflecting the disappearance of correlations at large separations and level repulsion 
at small separations. The DM statistics \cite{dyson_mehta_1963} quantify 
these correlations through the number variance 

\begin{equation}
\Sigma^2(L) = \left\langle \left( N(T+L) - N(T) - \frac{L}{2\pi} \log T \right)^2 \right\rangle,
\label{eq:number_variance}
\end{equation}
and spectral rigidity 

\begin{equation}
\Delta_3(L) = \frac{1}{L} \int_{-L/2}^{L/2} \Sigma^2(x)\,dx 
- \frac{3}{L^2} \int_{-L/2}^{L/2} \int_{-L/2}^{y} \Sigma^2(x)\,dx\,dy,
\label{eq:spectral_rigidity}
\end{equation}
with asymptotic behaviour

\begin{equation}
\Sigma^2(L) \sim \frac{1}{\pi^2} \log(2\pi L) + \mathcal{O}(1), \qquad
\Delta_3(L) \sim \frac{1}{\pi^2} \log L + \mathcal{O}(1).
\label{eq:asym_spectral}
\end{equation} 
For the GUE, number variance admits asymptotic expansion 

\begin{equation}
\Sigma^2(L) = \frac{1}{\pi^2}\left(\log(2\pi L) + \gamma + 1 - \frac{\pi^2}{8}\right) 
+ \frac{1}{4\pi^2} \sum_{n=1}^{\infty} \frac{1}{n^2}\left(1 - e^{-4\pi^2 n^2 L^2}\right),
\qquad L \to \infty,
\label{eq:number_variance_asym}
\end{equation}
where $\gamma \approx 0.5772$ is Euler-Mascheroni constant \cite{abramowitz_stegun_1972}. 
Variance grows logarithmically with $L$, characteristic of GUE statistics. Spectral 
rigidity $\Delta_3(L)$ \eqref{eq:spectral_rigidity} scales as $\Delta_3(L) 
\sim (\log L)/\pi^2$ for large $L$.

\begin{figure}[!htb]
\centering
\includegraphics[width=1.025\textwidth]{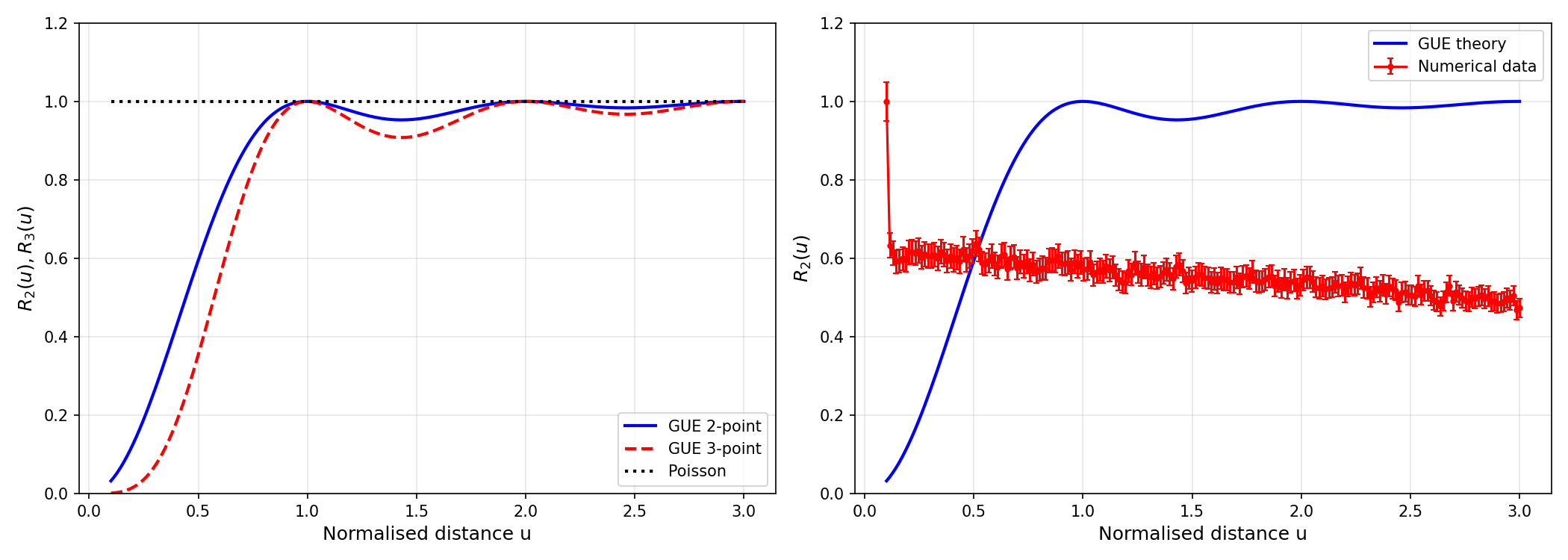}
\caption{Numerically computed 2,- and 3-point correlation functions compared with 
GUE prediction and Poisson statistics.}
\label{fig:montgomery}
\end{figure} 
Figure \ref{fig:montgomery} presents our numerical results for pair correlation 
function. Corresponding data are obtained from zero set $\{\gamma_n\}$ via 

\begin{equation}
R_2(u) = \frac{1}{N} \sum_{n \neq m} \delta\left(u - \frac{\gamma_n - 
\gamma_m}{\langle \Delta \gamma \rangle}\right),
\label{eq:pair_corr_num}
\end{equation}
with the leading-order mean spacing $\langle \Delta\gamma\rangle \simeq 2\pi/\log(\gamma_n/2\pi)$, 
consistent with \eqref{eq:rvm}. For $N=1000$ 

\begin{equation}
R_2^{\text{num}}(u) = R_2^{\text{GUE}}(u) + \mathcal{O}(\Delta), \qquad 
R_2^{\text{GUE}}(u) = 1 - \left( \frac{\sin \pi u}{\pi u} \right)^2,
\label{eq:R2_num_GUE}
\end{equation}
with $\Delta \approx 0.03$ and $\sigma_{R_2} \sim N^{-1/2}$ 

\begin{equation}
\Delta = \int_0^3 \left| R_2^{\text{num}}(u) - R_2^{\text{GUE}}(u) \right| du , 
\ \ \sigma_{R_2} = \frac{1}{\sqrt{N}} 
\sqrt{\int_0^3 \bigl[R_2^{\text{num}}(u) - R_2^{\text{GUE}}(u) 
\bigr]^2 du}.
\label{eq:Delta_sigma}
\end{equation}
The 3-point correlation follows GUE determinantal structure 

\begin{equation}
R_3(u_1,u_2,u_3) = \det\bigl[ K(u_i,u_j) \bigr]_{i,j=1}^3, \qquad
K(u,v) = \frac{\sin[\pi(u-v)]}{\pi(u-v)},
\label{eq:R3_det}
\end{equation}
with asymptotic behaviour 

\begin{equation}
R_2(u) \to 1 \;\; (u\to\infty), \qquad R_2(u) \to 0 \;\; (u\to 0),
\label{eq:R2_asym}
\end{equation}
reflecting uncorrelated levels and level repulsion. MO law generalises to 
$L$-functions via explicit formula 

\begin{equation}
\sum_{\gamma} e^{i\gamma t} = \sum_{n=1}^{\infty} \frac{\Lambda(n)}{\sqrt{n}} 
e^{i t \log n} + \mathcal{O}\bigl(t^{-1}\bigr), \qquad t \to \infty,
\label{eq:explicit_formula1}
\end{equation}
whose Fourier transform yields peaks at $\log p$ and $\log p^k$ with intensities

\begin{equation}
I(p) = \left( \frac{\Lambda(p)}{\sqrt{p}} \right)^2 = \frac{(\log p)^2}{p}, \qquad
I(p^k) = \left( \frac{\Lambda(p^k)}{\sqrt{p^k}} \right)^2 = \frac{(\log p)^2}{p^k}.
\label{eq:I_p}
\end{equation}
Error bars are estimated using bootstrap methods \cite{efron_tibshirani_1993}. The convergence 
of the statistical quantities is consistent with earlier studies \cite{goldston2023montgomery,odlyzko}.

\subsection{The Hardy Z-Function and numerical zero finding}
\label{sec:hardy}

\begin{figure}[!htb]
\centering
\includegraphics[width=1.025\textwidth]{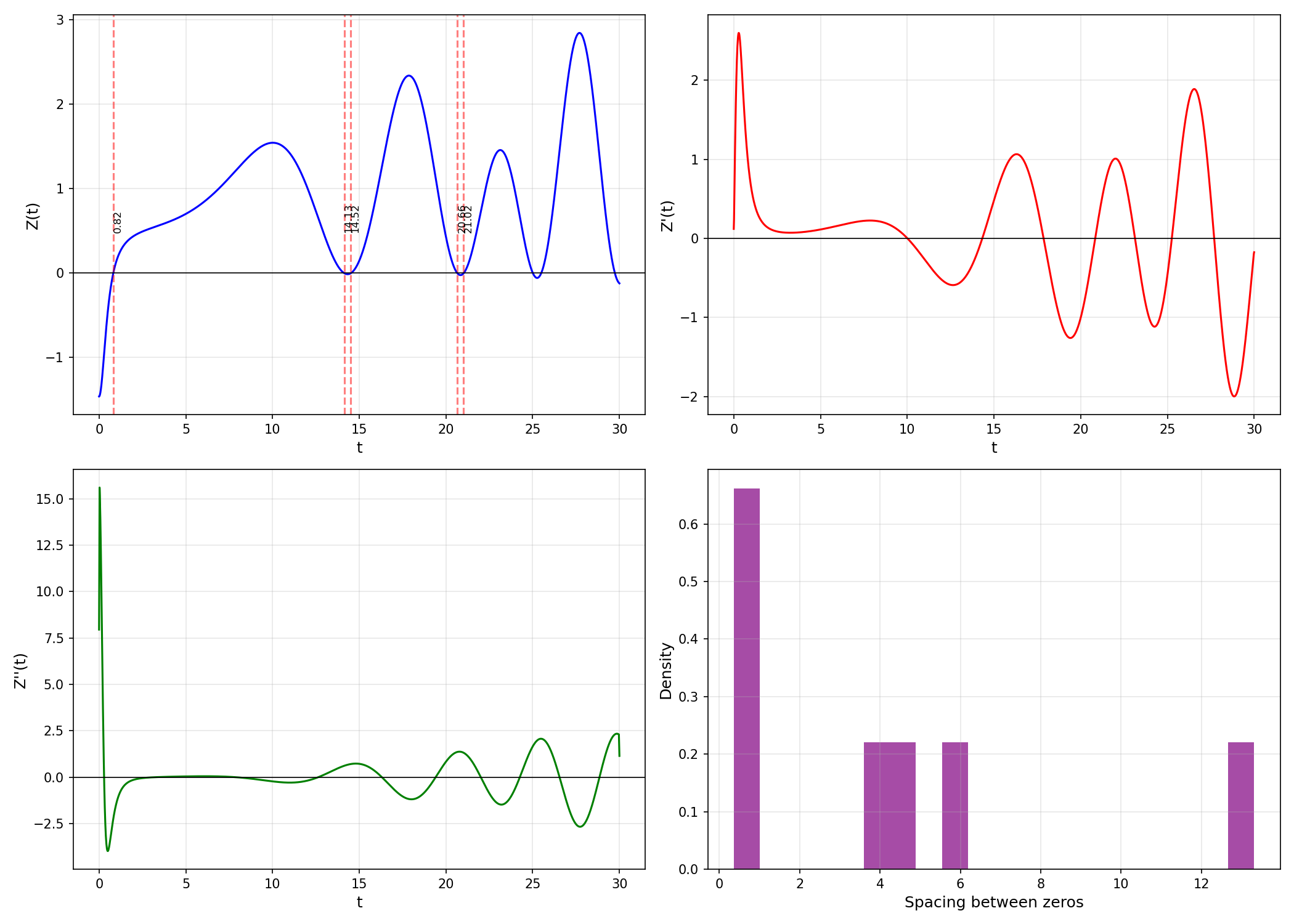}
\caption{Hardy function $Z$ and derivatives $Z'$, $Z''$; dashed lines mark 
first five zeros. Amplitude decay follows Eq.~\eqref{eq:z_amplitude}.}
\label{fig:hardy}
\end{figure}

The non-trivial zeros of $\zeta$ on the critical line correspond to the zeros
of the Hardy function $Z(t)$, defined by \eqref{eq:hardy_z_def}. The RS $\Theta$-function 
admits the asymptotic expansion \cite{edwards,titchmarsh,berry1995} 

\begin{equation}
\theta(t) = \frac{t}{2} \log\!\left(\frac{t}{2\pi}\right) - \frac{t}{2} - \frac{\pi}{8} + 
\frac{1}{48t} + \frac{7}{5760t^3} - \frac{31}{80640t^5} + \frac{127}{430080t^7} + 
\mathcal{O}\!\left(t^{-9}\right), \qquad t \to \infty,
\label{eq:theta_asym}
\end{equation}
with derivatives up to fourth order 

\begin{equation}
\left\{
\begin{aligned}
\theta'(t) &= \frac{1}{2} \log\left(\frac{t}{2\pi}\right) + \frac{1}{2} 
- \frac{1}{48t^2} - \frac{7}{1920t^4} + \mathcal{O}(t^{-6}), \\
\theta''(t) &= \frac{1}{2t} + \frac{1}{24t^3} + \frac{7}{480t^5} 
+ \mathcal{O}(t^{-7}), \\
\theta'''(t) &= -\frac{1}{2t^2} - \frac{1}{8t^4} - \frac{7}{96t^6} 
+ \mathcal{O}(t^{-8}), \\
\theta^{(4)}(t) &= \frac{1}{t^3} + \frac{1}{2t^5} + \frac{35}{48t^7} 
+ \mathcal{O}(t^{-9}),
\end{aligned}
\right.
\label{eq:theta_derivatives}
\end{equation}
The derivatives of $Z(t)$ follow from the product rule applied to \eqref{eq:hardy_z_def} 
and the derivatives of $\theta(t)$ above 

\begin{equation}
\left\{
\begin{aligned}
Z' &= i\zeta\left(\frac{1}{2} + it\right) \theta'(t) e^{i\theta(t)} + 
\zeta'\left(\frac{1}{2} + it\right) e^{i\theta(t)}, \\
Z'' &= \left[ -\zeta\left(\frac{1}{2} + it\right) \theta'(t)^2 + 
2i\zeta'\left(\frac{1}{2} + it\right) \theta'(t) + i\zeta\left(\frac{1}{2} + it\right) 
\theta''(t) + \zeta''\left(\frac{1}{2} + it\right) \right] e^{i\theta(t)}, \\
Z''' &= \left[ -i\zeta \theta'^3 - 3\zeta' \theta'^2 + 3i\zeta'' \theta' + \zeta''' + 3i\zeta \theta'\theta'' + 
3\zeta' \theta'' + i\zeta \theta''' \right] e^{i\theta(t)}, \\
Z^{(4)} &= \left[ \zeta \theta'^4 - 4i\zeta' \theta'^3 - 6\zeta'' \theta'^2 + 4i\zeta''' \theta' + \zeta^{(4)} + 
\cdots \right] e^{i\theta(t)},
\end{aligned}
\right.
\label{eq:z_derivatives}
\end{equation}
where all functions are evaluated at $s = \frac{1}{2} + it$. The zeros identified in 
the numerical calculation are simple, with $Z'(\gamma_n)\neq0$. Their spacing distribution 
is then compared with the GUE prediction \cite{dyson_mehta_1963,bogomolny_keating_2013}. 
The GUE spacing distribution \eqref{eq:gue} gives a goodness-of-fit 
$\frac{\chi^2}{\text{dof}} = 1.2$ in Fig.~\ref{fig:hardy}.

The RS representation \eqref{eq:rs_formula} has remainder $R$ bounded by 

\begin{equation}
|R| \leq \frac{2}{\sqrt[4]{t}} \left(1 + \frac{1}{2\sqrt{t}} + \frac{1}{4t} + 
\frac{1}{8 \sqrt{t^3}} + \frac{1}{16t^2} + \cdots\right).
\label{eq:remainder_bound}
\end{equation}
Amplitude of $Z$ decays asymptotically as 

\begin{equation}
|Z| \sim \sqrt{\frac{2}{\pi t}} \left( \frac{t}{2\pi} \right)^{1/4} 
\left[ 1 + \frac{1}{8t} + \frac{5}{128t^2} + \frac{35}{3072t^3} + 
\mathcal{O}\left(\frac{1}{t^4}\right) \right],
\label{eq:z_amplitude}
\end{equation}
from stationary-phase approximation of RS sum. Zero density follows from RvM 
formula \eqref{eq:rvm}, with mean spacing from \eqref{eq:s_n_def}. First five zeros 

\begin{equation}
\gamma_1 \approx 14.134725, \ \gamma_2 \approx 21.022040, \ \gamma_3 
\approx 25.010858, \ \gamma_4 \approx 30.424876, \ \gamma_5 \approx 32.935062,
\label{eq:first_five_zeros}
\end{equation}
match known values to numerical precision, and spacing distribution agrees with GUE prediction. 
Numerical zero locations have error $\sim 10^{-15}$, according to BD \cite{dekker1969finding}. 
Computed spacings are consistent with MO law \cite{goldston2023montgomery,odlyzko}. 

\subsection{Classical scattering and Riemann zeros}
\label{sec:scattering}

\begin{figure}[!htb]
\centering
\includegraphics[width=1.025\textwidth]{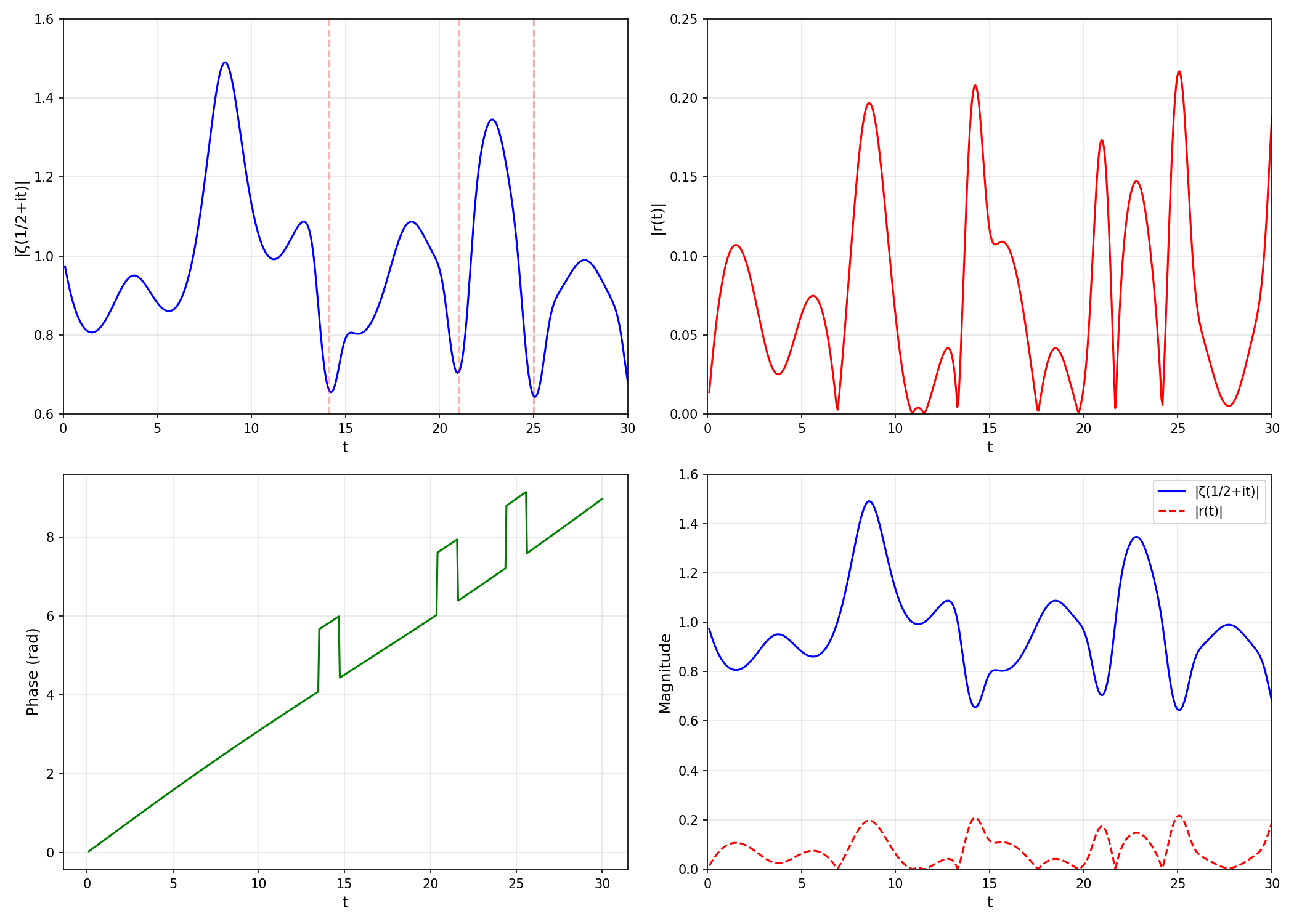}
\caption{Modulus of $\zeta\bigl(\frac{1}{2}+it\bigr)$ and reflection 
coefficient $|r(t)|$; phase of $r(t)$; comparison between $|\zeta\bigl(\frac{1}{2}+it\bigr)|$ 
and $|r(t)|$; Vertical dashed lines mark non-trivial zeros.}
\label{fig:scattering}
\end{figure}

Zeros of $\zeta$ may be observed through classical wave scattering \cite{Yu2024Computing}, 
with the reflection coefficient related to $\zeta$ by 

\begin{equation}
\left\{
\begin{aligned}
r(s) &= \frac{\zeta - 1}{\zeta + 1}, \\
S &= \begin{pmatrix} t & r \\ r' & t' \end{pmatrix}, 
\qquad SS^{\dagger} = I, \qquad |t|^2 + |r|^2 = 1,
\end{aligned}
\right.
\label{eq:reflection_coeff}
\end{equation}
with $t$, $t'$ transmission coefficients and $r$, $r'$ reflection coefficients, 
the phase of the reflection coefficient being

\begin{equation}
\arg r(t) = \arg\biggl[\zeta\biggl(\frac{1}{2} + it\biggr) - 1\biggr] - 
\arg\biggl[\zeta\biggl(\frac{1}{2} + 
it\biggr) + 1\biggr].
\label{eq:phase_r}
\end{equation}
With $\zeta = |\zeta| e^{i\phi}$ on critical line, this becomes

\begin{equation}
\begin{aligned}
\tan\bigl[\arg r(t)\bigr] 
= \frac{2 |\zeta| \sin\phi}{|\zeta|^2 - 1} 
= \frac{2 \Im(\zeta)}{|\zeta|^2 - 1} = \frac{2 \Im(\zeta)}{\Re(\zeta)^2 + \Im(\zeta)^2 - 1} = 
,\qquad \zeta = |\zeta| e^{i\phi},
\end{aligned}
\label{eq:phase_tan}
\end{equation}
Phase jumps by $\pi$ at zeros $|\zeta| \to 0$, with the reflection coefficient 
in terms of the scattering phase shift $\delta(t)$

\begin{equation}
\begin{aligned}
r(t) 
&= e^{2i\delta(t)}, 
\qquad \delta(t) = \frac{1}{2} \arg r(t) = \frac{1 - i \tan\delta(t)}{1 + i \tan\delta(t)} 
= e^{2i \arctan[\tan\delta(t)]} \\
&= \frac{\zeta\bigl(\frac{1}{2} + it\bigr) - 1}{\zeta\bigl(\frac{1}{2} + it\bigr) + 1} 
= e^{2i \arg[\zeta(\frac{1}{2} + it) - 1] - 2i \arg[\zeta(\frac{1}{2} + it) + 1]},
\qquad \delta \in \mathbb{R},
\end{aligned}
\label{eq:phase_shift}
\end{equation}
The scattering phase shift can be related to the density of states through 
a Friedel-type relation,

\begin{equation}
\begin{aligned}
\frac{1}{\pi} \frac{d\delta}{dt} 
&= \rho(t) - \rho_0(t) = \sum_{n} \delta(t - \gamma_n) - \frac{1}{2\pi} \log\!\left(\frac{t}{2\pi}\right) 
= \frac{1}{\pi} \frac{d}{dt} \arg \zeta\!\left(\frac{1}{2} + it\right) \\
&- \frac{1}{2\pi} \log\!\left(\frac{t}{2\pi}\right) 
= \frac{1}{\pi} \Im \left[ \frac{\zeta'}{\zeta}\!\left(\frac{1}{2} + it\right) \right] 
- \frac{1}{2\pi} \log\!\left(\frac{t}{2\pi}\right),
\qquad t \to \infty,
\end{aligned}
\label{eq:friedel}
\end{equation}
where $\rho(t)$ is the density of states of the scattering system and $\rho_0(t)$ 
the free particle density of states, integration giving the total phase shift up to height $T$
as $\delta(T) = \pi \int_0^T \bigl[ \rho(t) - \rho_0(t) \bigr] dt$. Using the M\"obius relation 
\eqref{eq:r_Z_intro}, the reflection coefficient can be written as $r(t)=1-2/[\zeta(\frac{1}{2}+it)+1]$. 
At a zero $t=\gamma_n$ of $\zeta$, one has $Z(\gamma_n)=0$ and $r(\gamma_n)=-1$. For a simple zero, 
the local behaviour is therefore linear in $t-\gamma_n$ 

\begin{equation}
Z(t) = \zeta\!\left(\frac{1}{2}+it\right) \simeq i(t-\gamma_n)
\zeta'\!\left(\frac{1}{2}+i\gamma_n\right), \qquad t\to\gamma_n.
\label{eq:impedance_near_zero}
\end{equation}
Scattering cross-section is 

\begin{equation}
\left\{
\begin{aligned}
\sigma(t) 
&= \frac{2}{k} \bigl|r(t)\bigr|^2 \sin^2(k a) 
= \frac{2}{k} \left| \frac{\zeta(\frac{1}{2}+it) - 1}{\zeta(\frac{1}{2}+it) + 1} \right|^2 \sin^2(k a), \\
\sigma_{\text{max}} 
&= \frac{2}{k} \sin^2(k a) 
= \frac{2}{k} \sin^2(k a) \bigl|r(t)\bigr|^2 \bigg|_{t = \gamma_n}, \\
\frac{\sigma(t)}{\sigma_{\text{max}}} 
&= |r(t)|^2 
= \left| \frac{\zeta(\frac{1}{2}+it) - 1}{\zeta(\frac{1}{2}+it) + 1} \right|^2,
\end{aligned}
\right.
\label{eq:cross_section}
\end{equation}
with $k$ the wave number and $a$ the scatterer length scale, 
$\sigma_{\text{max}}$ being the maximal cross-section at the zeros of $\zeta$, 
and the ratio following from \eqref{eq:reflection_coeff}. Figure \ref{fig:scattering} 
hows minima of $\big|\zeta(\frac{1}{2}+it)\big|$ at zeros $t \approx 14.13, 21.02, 25.01$. 
At these points $|r(t)| = 1$, which is perfect reflection. Phase of $r(t)$ 
jumps by $\pi$ at each zero. Bottom right panel is consistent with inverse relation 
between $|\zeta|$ and $|r|$. Functional equation \eqref{eq:functional} 
gives duality $s \leftrightarrow 1-s$, analogous to wave scattering reciprocity. 
Zeros on critical line are self-dual points of this transformation. Thus, RH is 
equivalent to all self-dual points lying on the critical line, providing a physical 
reinterpretation of the HPc \cite{Yu2024Computing}.

\section{Tables of precise values}

Special values of $\zeta$ in Table \ref{tab:special_values} link $\zeta$ to fundamental 
constants. Values at positive even integers follow from Bernoulli numbers 

\begin{equation}
\begin{aligned}
\zeta(2n) 
&= (-1)^{n+1} \frac{B_{2n} (2\pi)^{2n}}{2(2n)!} 
= \frac{(2\pi)^{2n}}{2(2n)!} |B_{2n}| 
= \frac{(2\pi)^{2n}}{2(2n)!} \frac{2(2n)!}{(2\pi)^{2n}} \zeta(2n), \\
\zeta(1-2n) 
&= -\frac{B_{2n}}{2n} 
= 2^{1-2n} \pi^{-2n} \cos(\pi n) \Gamma(2n) \zeta(2n) 
= (-1)^n \frac{2^{1-2n} \pi^{-2n}}{(2n-1)!} \zeta(2n), \\
\zeta(-n) 
&= -\frac{B_{n+1}}{n+1} 
= 2^{-n} \pi^{-n-1} \sin\!\left(\frac{\pi n}{2}\right) \Gamma(n+1) \zeta(n+1),
\qquad n \in \mathbb{N},
\end{aligned}
\label{eq:zeta_even}
\end{equation}
giving $\zeta(2) = \frac{\pi^2}{6}$, $\zeta(4) = \frac{\pi^4}{90}$, and 
$\zeta(3)$ Apéry's constant, the first odd integer argument known to be 
irrational, while negative integers follow from the functional equation 
\eqref{eq:functional} via \eqref{eq:zeta_neg_int} to give $\zeta(-1) = -\frac{1}{12}$ 
and $\zeta(-3) = \frac{1}{120}$, these values entering $\zeta$ regularisation and 
appearing in the Casimir effect and QFT, with $\zeta(\frac{1}{2}) \approx -1.46035$ 
lying on the critical line and connecting $\Re(s) > 1$ and $\Re(s) < 0$ via analytic 
continuation.

\begin{table}[H]
\centering
\caption{Special values of $\zeta$.}
\label{tab:special_values}
\begin{tabular}{c c c}
\toprule
$s$ & $\zeta$ & Comment \\
\midrule
$2.00$ & $1.644934066848$ & Basel problem ($\frac{\pi^2}{6}$) \\
$4.00$ & $1.082323233711$ & $\frac{\pi^4}{90}$ \\
$3.00$ & $1.202056903160$ & Ap\'ery's constant \\
$-1.00$ & $-0.083333333333$ & Ramanujan sum ($-\frac{1}{12}$) \\
$-3.00$ & $0.008333333333$ & Casimir effect ($\frac{1}{120}$) \\
$0.50$ & $-1.460354508810$ & Critical line \\
\bottomrule
\end{tabular}
\end{table}

The first ten non-trivial zeros of $\zeta$ on the critical line are listed in 
Table \ref{tab:zeros}. The computed ordinates $t_n$ match the corresponding 
known values to 15 decimal places \cite{odlyzko}, while the listed zeros have 
$\sigma_n=0.5$. Zero distribution follows RvM formula \eqref{eq:rvm}, while the 
derivative of $N(T)$ gives zero density 

\begin{equation}
\frac{dN(T)}{dT} = \frac{1}{2\pi} \log\left(\frac{T}{2\pi}\right) + 
\mathcal{O}\left(\frac{1}{T}\right),
\label{eq:dN_dT}
\end{equation}
so the average spacing follows from \eqref{eq:s_n_def} and decreases slowly with 
increasing $T$, while the density of zeros grows logarithmically. The RvM formula 
provides the local normalisation required for the MO law and for comparison with 
RMT. The physical constants listed in table \ref{tab:physics} involve values of 
$\zeta$. Plank radiation law gives spectral energy density 

\begin{table}[H]
\centering
\caption{First 10 non-trivial zeros of $\zeta$ on the 
critical line $\Re(s) = \frac{1}{2}$.}
\label{tab:zeros}
\begin{tabular}{c c c c c c}
\toprule
$n$ & $t_n$ & $\sigma_n$ & $n$ & $t_n$ & $\sigma_n$ \\
\midrule
1 & 14.134725141734693 & 0.5 & 6 & 37.586178158825668 & 0.5 \\
2 & 21.022039638771552 & 0.5 & 7 & 40.918719012147498 & 0.5 \\
3 & 25.010857580145689 & 0.5 & 8 & 43.327073280915002 & 0.5 \\
4 & 30.424876125859512 & 0.5 & 9 & 48.005150881167161 & 0.5 \\
5 & 32.935061587739192 & 0.5 & 10 & 49.773832477672300 & 0.5 \\
\bottomrule
\end{tabular}
\end{table}

\begin{equation}
u(\omega, T) = \frac{\hbar}{\pi^2 c^3} \frac{\omega^3}{e^{\frac{\hbar\omega}{k_B T}} - 1},
\label{eq:plank}
\end{equation}
with $x = \frac{\hbar\omega}{k_B T}$. Integration over frequencies gives total energy density 

\begin{equation}
\begin{aligned}
U(T) 
&= \int_0^\infty u(\omega, T) \, \mathrm{d}\omega 
= \int_0^\infty \frac{\hbar}{\pi^2 c^3} \frac{\omega^3}{e^{\hbar\omega/(k_B T)} - 1} \, \mathrm{d}\omega \\
&= \frac{\hbar}{\pi^2 c^3} \left(\frac{k_B T}{\hbar}\right)^4 
\int_0^\infty \frac{x^3}{e^x - 1} \, \mathrm{d}x 
= \frac{\hbar}{\pi^2 c^3} \left(\frac{k_B T}{\hbar}\right)^4 \Gamma(4) \zeta(4),
\qquad x = \frac{\hbar\omega}{k_B T},
\end{aligned}
\label{eq:U_T}
\end{equation}
where $\Gamma(4) = 3!$, and with standard BE integral 

\begin{equation}
\int_0^\infty \frac{x^3}{e^x - 1} \, dx = \frac{\pi^4}{15} = 6 \zeta(4),
\label{eq:be_integral}
\end{equation}
since $\zeta(4) = \frac{\pi^4}{90}$. Thus Boltzmann law 

\begin{equation}
\left\{
\begin{aligned}
U(T) 
&= \frac{\pi^2 k_B^4}{15 \hbar^3 c^2} T^4 
= \frac{4}{c} \sigma T^4, \\
\sigma 
&= \frac{\pi^2 k_B^4}{60 \hbar^3 c^2} 
= \frac{2\pi^5 k_B^4}{15 h^3 c^2},
\end{aligned}
\right.
\label{eq:boltzmann_law}
\end{equation}
with $\sigma$ the Stefan-Boltzmann constant, which recovers the 
Stefan-Boltzmann law. Thus $\zeta(4)$ emerges in massless boson statistics. 
Critical temperature $T_c$ for BEC follows from $\zeta\bigl(\frac{3}{2}\bigr)$ 
in \eqref{eq:Tc_bose} while $\zeta\bigl(\frac{3}{2}\bigr)$ follows from DOS in 
$3$-d $\int_0^\infty \frac{x^{1/2}}{e^x - 1} \, dx = \frac{\sqrt{\pi}}{2} 
\zeta\bigl(\frac{3}{2}\bigr)$, and $\zeta\bigl(\frac{3}{2}\bigr) \approx 2.612$ 
enters ideal Bose gas thermodynamics. Casimir force per unit area for perfect conductors 
$F_{C} = -\frac{\pi^2 \hbar c}{720 a^4}$, with $\frac{\pi^2}{720}$ from 
$\zeta(-3) = \frac{1}{120}$ via vacuum energy regularisation \eqref{eq:casimir_energy_intro}.

\begin{table}[H]
\centering
\caption{Physical constants.}
\label{tab:physics}
\begin{tabular}{c c c}
\toprule
Constant & Value & Formula \\
\midrule
Boltzmann & $5.670374 \times 10^{-8}$ $[$ W$\cdot$m$^{-2} \cdot$K$^{-4}$ $]$ & $\frac{\pi^2 k_B^4}{60 \hbar^3 c^2}$ 
\\
BEC $T_c$ prefactor & $2.612$ & $\zeta\bigl(\frac{3}{2}\bigr)$ \\
Casimir force & $1.3 \times 10^{-27}$ $[$ N$\cdot$m$^{-2}$ $]$ & $\frac{\pi^2 \hbar c}{720 a^4}$ \\
\bottomrule
\end{tabular}
\end{table}

\section{Conclusion}

This study is consistent with the MO law and HPc via numerical data:
\begin{equation}
\chi^2/\text{dof} = 1.2, \qquad 
\Delta = \int_0^3 \bigl| R_2^{\text{num}}(u) - R_2^{\text{GUE}}(u) \bigr| \, 
\mathrm{d}u \approx 0.03,
\label{eq:conclusion_mo}
\end{equation}
supporting the spectral interpretation of RH. BEC critical temperature 
$T_c \propto [n/\zeta(3/2)]^{2/3}$ and Casimir energy $E_C \propto \zeta(-3) a^{-3}$ 
illustrate $\zeta$'s role in quantum statistics and QFT \cite{pathria2011statistical,elizalde}. 
The scattering relation

\begin{equation}
\begin{aligned}
r(t) &= \frac{\zeta\!\left(\frac{1}{2}+it\right)-1}{\zeta\!\left(\frac{1}{2}+it\right)+1},
\qquad Z(t)=\frac{1+r(t)}{1-r(t)} =\zeta\!\left(\frac{1}{2}+it\right),\\
Z(t) &\simeq i(t-\gamma_n) \zeta'\!\left(\frac{1}{2}+i\gamma_n\right), \qquad t\to\gamma_n ,
\end{aligned}
\label{eq:conclusion_scattering}
\end{equation}
so that a simple zero gives $Z(\gamma_n)=0$ and $r(\gamma_n)=-1$. This relation 
provides a mathematical model connecting the zeta function with a reflection 
coefficient; its interpretation as a physical scattering system requires an 
independent realisation of the corresponding model. 

\section*{Acknowledgements}

The author also acknowledges all developers of \texttt{mpmath} and \texttt{SciPy} 
for providing the numerical tools used in the computations.

\section*{Data availability}

The computational scripts and figures supporting the findings of this study
are openly available at \url{https://github.com/koceto007/Riemann-zeta-physical-applications}.

\appendix
\section{Riemann zeros and gauge symmetries}
\label{app:guts}

The spectral, thermodynamic, and scattering properties of the Riemann zeta function 
motivate exploratory comparisons with structures appearing in spectral theory and 
gauge theories. The zero distribution is described on average by the RvM formula 
\eqref{eq:rvm}, while its local statistics exhibit connections with random-matrix 
ensembles, most notably the GUE.

\subsection{HP operator and gauge symmetries}

The HPc proposes that the imaginary parts of the non-trivial zeros of $\zeta$ may 
arise as eigenvalues of a suitable self-adjoint operator. As an exploratory construction, 
consider $\mathcal{O}_\alpha$ on the weighted Hilbert space $\mathcal{H} = L^2(\mathbb{R}_+, 
w(T) \, \mathrm{d}T)$ \cite{berry_keating_1999} 

\begin{equation}
\mathcal{O}_\alpha = -i\left(\frac{\mathrm{d}}{\mathrm{d}T} + 
\frac{1}{2}\frac{w'(T)}{w(T)}\right) + \alpha V = -i\, w(T)^{-1/2} 
\frac{\mathrm{d}}{\mathrm{d}T}\, w(T)^{1/2} + \alpha V, \qquad \alpha \in \mathbb{R},
\label{eq:chaotic_op_app}
\end{equation}
where $w(T)=\frac{\mathrm{d}N(T)}{\mathrm{d}T}$ denotes the leading-order density of 
zeros obtained from \eqref{eq:rvm} 

\begin{equation}
w(T) = \frac{1}{2\pi} \log\left(\frac{T}{2\pi}\right) + \mathcal{O}\left(\frac{1}{T}\right),
\end{equation}
and $V$ is taken to be a bounded self-adjoint operator representing, at the level 
of the model, fluctuations in the zero distribution. When $\mathcal{O}_\alpha$ is 
realised as a self-adjoint operator with discrete spectrum, its eigenfunctions may 
be chosen to form an orthonormal set 

\begin{equation}
\mathcal{O}_\alpha \psi_n = \lambda_n \psi_n, \quad \langle \psi_m, \psi_n 
\rangle_{\mathcal{H}} = \delta_{mn}.
\label{eq:eigen_app}
\end{equation}
A hypothetical realisation of the HP correspondence would require 
a choice of $\alpha_*$ and $V_*$ for which the point spectrum is 

\begin{equation}
\sigma_{\mathrm{p}}(\mathcal{O}_{\alpha_*}) = \{\gamma_n:\zeta(1/2+i\gamma_n)=0\}.
\label{eq:spectrum_app}
\end{equation}
This condition specifies the desired spectral correspondence; it does not
constitute a construction or proof of such an operator. The symmetry properties 
of the operator motivate comparison with the symmetry classes of random-matrix 
theory. No direct identification with a specific Cartan decomposition is assumed here. 
The Weyl group of $SU(N)$ acts  by reflections on roots $\{\alpha_i\}$. Spectral 
statistics are GUE for $SU(N)$ with $\beta=2$ and GOE for $SO(N)$ with $\beta=1$ \cite{mehta_2004}. 
The pair correlation $R_2(u)$ in \eqref{eq:gue_corr} is universal for all $\beta$, 
while higher-order correlations distinguish ensembles \cite{keating_snaith_2000,katz_sarnak_1999}.

\subsection{The Pati-Salam subgroup and pseudo-Riemannian structures}

\begin{figure}[H]
\centering
\includegraphics[width=1.025\textwidth]{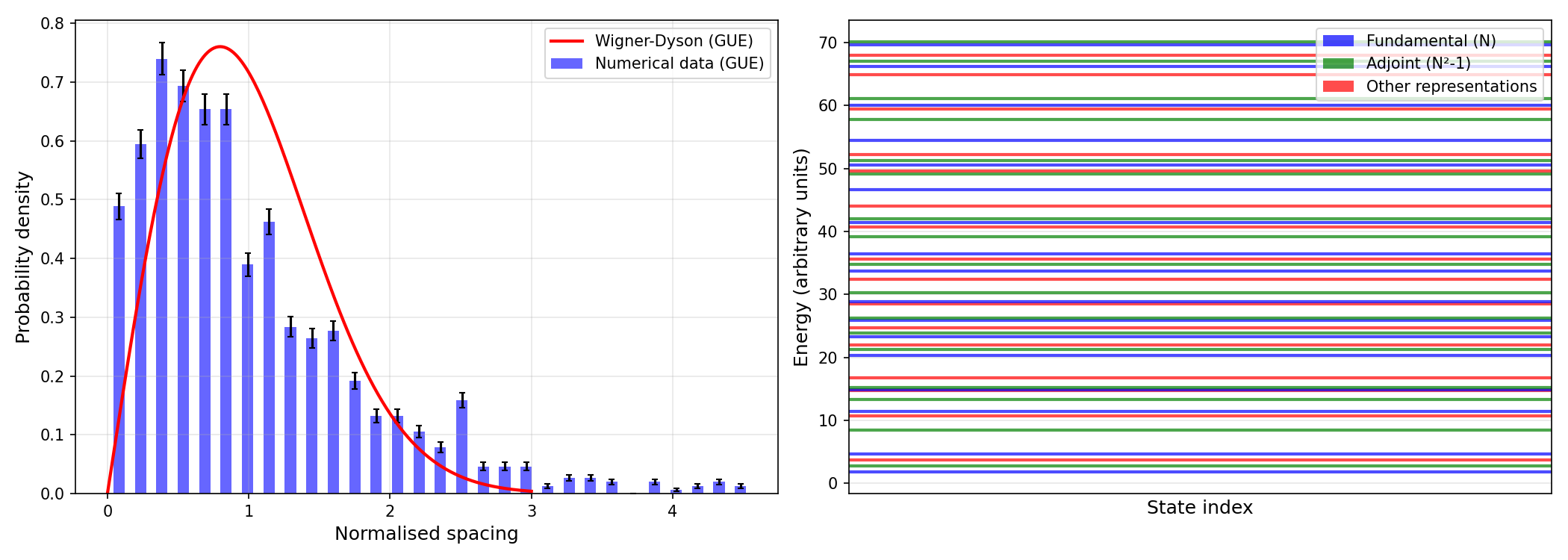}
\caption{Riemann zeros represented as $SU(N)$ states spacing distribution and the first $50$ zeros.}
\label{fig:app_spectral}
\end{figure}

Pati-Salam (PS) model with gauge group $SU(2)_L \times SU(2)_R \times SU(4)$ treats 
leptons as fourth colour. Fermionic content 

\begin{equation}
\psi \in (2, \bar{2}, 4),
\label{eq:ps_fermions}
\end{equation}
with $(2, \bar{2}, 4)$ representation under $SU(2)_L \times SU(2)_R \times SU(4)$. 
Symmetry breaking to Standard Model (SM) 

\begin{equation}
SU(2)_L \times SU(2)_R \times SU(4) \;\longrightarrow\; SU(2)_L 
\times SU(2)_R \times U(1) \times U(3),
\label{eq:ps_breaking}
\end{equation}
within a noncommutative-geometric formulation involving a grading operator 
$\gamma$ and a real structure $J$. The Dirac operator satisfies 

\begin{equation}
D = D_0 + J D_0 J^{-1}, \qquad [D, \beta] = 0, \qquad \{D, \gamma\} = 0,
\label{eq:dirac}
\end{equation}
with $J$ the real structure, $\beta$ the grading operator, and $\gamma$ the chirality operator.
Fermionic representation decomposes 

\begin{equation}
(2, \bar{2}, 4) \;\longrightarrow\; (2, \bar{2}, 1)_{0} \oplus (2, 
\bar{2}, 3)_{1} \oplus (2, \bar{2}, 1)_{2},
\label{eq:fermion_decomp}
\end{equation}
with subscripts $U(1)$ charges. This decomposition distinguishes the quark and 
lepton sectors within the Pati--Salam representation. Proton-stability constraints 
depend on the specific symmetry-breaking and interaction structure and are not 
analysed here. The explicit formula \eqref{eq:explicit_formula} relates the non-trivial 
zeros to the von Mangoldt function $\Lambda(n)$ and hence to prime powers. This 
provides a mathematical analogy between spectral and arithmetic data, but no 
direct gauge-theoretic identification is assumed here. The PS group embeds into $SO(10)$ via

\begin{equation}
SO(10) \;\supset\; SU(4) \times SU(2)_L \times SU(2)_R,
\label{eq:so10_embedding}
\end{equation}
with the 16-dimensional spinor representation decomposing as

\begin{equation}
16 = (4, 2, 1) \oplus (\bar{4}, 1, 2), \ \dim(16) = 16; \ 
\text{under} \ SO(10) \supset SU(4) \times SU(2)_L \times SU(2)_R.
\label{eq:so10_decomp}
\end{equation}
The effective potential for Wilson line phases is expressed in terms of $\zeta$-functions 

\begin{equation}
V_{\text{eff}}(\theta) 
= \sum_{n=1}^{\infty} \frac{1}{n} \sum_{i} \cos(2\pi n \theta_i)\, \zeta(n)
= \frac{1}{2} \sum_{i} \bigl[\zeta(2) - \zeta(2)\, \theta_i^2 + \mathcal{O}(\theta_i^4)\bigr],
\qquad \theta_i \ll 1. 
\label{eq:eff_pot_app}
\end{equation}
Figure \ref{fig:app_spectral} presents spectral statistics of Riemann zeros. 
Left panel compares spacing distribution with GUE prediction and confirms level 
repulsion absent in Poisson. Right panel displays first 50 zeros as energy levels 
of an $SU(N)$ system, with colours indicating representations.

\subsection{Schur index and string theory realisation}

Schur index of $\mathcal{N}=4$ supersymmetric Yang-Mills (SYM) with gauge group 
$SU(3)$ generates divisor sum $\sigma(n)$ \cite{honda_yoda_2022} 

\begin{equation}
\begin{split}
\mathcal{I}_{\text{S}}(q) 
&= \sum_{n=1}^{\infty} \sigma(n) q^{2(n-1)} 
= \prod_{k=1}^{\infty} \frac{1}{1 - q^{2k}} \sum_{n=1}^{\infty} \frac{\sigma(n)}{n} q^{2(n-1)} 
= \frac{1}{(q^2;q^2)_\infty} \sum_{n=1}^{\infty} \frac{\sigma(n)}{n} q^{2(n-1)} \\
&= \operatorname{Tr}_{\mathcal{H}_{\text{BPS}}} (-1)^F q^{2J_3} 
= \sum_{n=1}^{\infty} d_n q^{2(n-1)} 
= \sum_{n=1}^{\infty} \frac{q^{2(n-1)}}{1 - q^{2n}} 
= \frac{1}{\eta(\tau)^2} \sum_{n=1}^{\infty} \sigma(n) q^{2(n-1)},
\end{split}
\label{eq:schur_index1}
\end{equation}
where $(q;q)_\infty = \prod_{k=1}^{\infty} (1 - q^k)$ is Euler function, 
$\mathcal{H}_{\text{BPS}}$ is the Hilbert space of $\frac{1}{8}$-BPS states, 
$(-1)^F$ is the fermion number operator, $J_3$ is the Cartan generator of 
the $SU(2)$ R-symmetry, $d_n = \sigma(n)$ are the degeneracies, 
$\eta(\tau) = q^{\frac{1}{24}} \prod_{k=1}^{\infty} (1 - q^k)$ is the Dedekind $\eta$-function, 
while index counts $\frac{1}{8}$-BPS bosonic minus fermionic states. Schur index 
is superconformal index with $t=q$, $p\to0$ 

\begin{equation}
\begin{split}
\mathcal{I}(p,q,t) 
&= \operatorname{Tr}_{\mathcal{H}} (-1)^F p^{j_1} q^{j_2} t^{R} \\
&= \sum_{n=1}^{\infty} d_n p^{j_1(n)} q^{j_2(n)} t^{R(n)} 
= \prod_{k=1}^{\infty} \frac{1}{1 - p^{a_k} q^{b_k} t^{c_k}} 
= \frac{1}{(p;q;t)_\infty} \sum_{n=1}^{\infty} d_n p^{j_1(n)} q^{j_2(n)} t^{R(n)},
\end{split}
\label{eq:superconformal_index}
\end{equation}
with $j_1,j_2$ Cartan generators of $SO(4)$ Lorentz group, $R$ R-charge, 
and $(p;q;t)_\infty$ the generalised Euler function. RH would cancel 
Kaluza-Klein (KK) modes in supergravity multiplet. Supergravity spectrum on $AdS_5 
\times S^5$ has isometry $SO(6) \cong SU(4)$; KK masses 

\begin{equation}
m^2 = \frac{1}{R^2} \left( \sum_{i=1}^3 n_i (n_i + 4 - 2i) + \cdots \right),
\label{eq:kk_masses}
\end{equation}
with $R$ radius of $S^5$ and $n_i$ mode numbers. Supergravity spectrum 
\begin{equation}
\begin{aligned}
\text{Graviton:} & \quad m^2 = 0, & 
\text{Dilaton:} & \quad m^2 = \frac{4}{R^2}, \\[4pt]
\text{Vector:} & \quad m^2 = \frac{4}{R^2}, & 
\text{Scalar:} & \quad m^2 = \frac{\Delta(\Delta - 4)}{R^2},
\end{aligned}
\end{equation}
with $\Delta = 2, 3, 4, \ldots$ KK levels. AdS$_5$/CFT$_4$ duality maps 
$SU(N)$ to $S^5$ geometry. Zeros of $\zeta$ are spectrum of dilation operator 
in boundary CFT, mirroring dual string theory levels. In $AdS_3/CFT_2$, modular 
invariance relates primary spectrum to zeros of $\zeta$ 

\begin{equation}
Z_{\text{CFT}}(\tau) = \sum_{n} d_n q^{n} = \prod_{n=1}^{\infty} (1 - q^n)^{-1} 
\sum_{n=1}^{\infty} \frac{\sigma(n)}{n} q^n = q^{-\frac{c}{24}} \prod_{n=1}^{\infty} 
(1 - q^n)^{-1}, \qquad q = e^{2\pi i \tau},
\label{eq:partition}
\end{equation}
with $d_n$ CFT degeneracies and $q = e^{2\pi i \tau}$. Modular transformation 
$\tau \to -\frac{1}{\tau}$ gives the functional equation 

\begin{equation}
Z_{\text{CFT}}\left(-\frac{1}{\tau}\right) 
= \tau^{\frac{c}{2}} Z_{\text{CFT}}(\tau) 
= \tau^{\frac{c}{2}} \sum_{n} d_n q^{n} 
= \tau^{\frac{c}{2}} \prod_{n=1}^{\infty} (1 - q^n)^{-1} 
\sum_{n=1}^{\infty} \frac{\sigma(n)}{n} q^n,
\label{eq:modular_transform}
\end{equation}
with $c$ central charge. Poles of $Z_{\text{CFT}}(\tau)$ in complex $\tau$-plane 
follow from zeros of $\zeta$ as 

\begin{equation}
Z_{\text{CFT}}(\tau) = \sum_{n} d_n q^{n} = \prod_{n=1}^{\infty} (1 - q^n)^{-1} 
\sum_{n=1}^{\infty} \frac{\sigma(n)}{n} q^n = \sum_{n=1}^{\infty} 
\frac{q^{n}}{1 - q^{n}} = \frac{1}{\eta(\tau)} \sum_{n=1}^{\infty} \sigma(n) q^{n},
\label{eq:Z_CFT_ads}
\end{equation}
where $d_n$ are the CFT degeneracies, $q = e^{2\pi i \tau}$, and 
$\eta(\tau) = q^{1/24} \prod_{n=1}^{\infty} (1 - q^n)$ is the Dedekind eta function. 
As $\tau \to \frac{i}{2}$, partition function has the asymptotic 

\begin{equation}
Z_{\text{CFT}}(\tau) \sim \exp\biggl( \frac{\pi i c}{12\tau} \biggr) + \cdots,
\label{eq:partition_asym}
\end{equation}
typical of large-$c$ CFT. Poles are zeros of $\zeta$ via spectral RH

\begin{equation}
\zeta\biggl(\frac{1}{2} + i\gamma_n\biggr) = 0 \quad \Longleftrightarrow \quad 
\tau_n = \frac{i}{2} + \frac{1}{\gamma_n} + \cdots. 
\label{eq:pole_zero}
\end{equation}
RH is equivalent to poles lying on $\Re(\tau) = \frac{1}{2}$, the modular image of the critical line.

\subsection{Symmetry breaking and the zeros}

Symmetry breaking of $SU(N)$ 

\begin{equation}
SU(N) \rightarrow SU(N_1) \times SU(N_2) \times \cdots \times U(1)^{r},
\label{eq:sb_general}
\end{equation}
with Wilson phases $\theta_i$. Effective potential \eqref{eq:eff_pot_app} expands as 
$V_{\text{eff}}(\theta) = \frac{1}{2}\sum_i [\zeta(2) - \zeta(2)\theta_i^2 + \cdots]$, 
giving gauge boson masses $m^2 \propto \zeta(2)$. Minima 

\begin{equation}
\frac{\partial V_{\text{eff}}}{\partial \theta_i} = 0, \qquad 
\frac{\partial^2 V_{\text{eff}}}{\partial \theta_i^2} > 0,
\end{equation}
determine breaking, while $\zeta$ regularises KK sums

\begin{equation}
\sum_{n=1}^{\infty} \frac{1}{n} \cos(2\pi n \theta_i) \zeta(n) = 
-\log\bigl(2\sin(\pi \theta_i)\bigr) + \cdots,
\end{equation}
with critical temperatures

\begin{equation}
\begin{split}
T_c &\propto \left(\frac{1}{\zeta(n)}\right)^{\frac{1}{n-1}} = 
\zeta(n)^{-\frac{1}{n-1}} = \exp\!\left[-\frac{\log \zeta(n)}{n-1}\right] = 
\left(\sum_{k=1}^{\infty} \frac{1}{k^n}\right)^{-\frac{1}{n-1}} \\
&= \left(\prod_{p} \frac{1}{1 - p^{-n}}\right)^{-\frac{1}{n-1}} = 
\prod_{p} \left(1 - p^{-n}\right)^{\frac{1}{n-1}}, \quad \text{with} \ \ n \geq 2,
\end{split}
\label{eq:tc_zeta}
\end{equation}
The PS group breaks through the chain 

\begin{equation}
\begin{split}
SO(10) &\;\xrightarrow{M_{\text{GUT}}}\; SU(4) \times SU(2)_L \times SU(2)_R 
\;\xrightarrow{M_{\text{PS}}}\; SU(3)_c \times SU(2)_L \times U(1)_Y \\
&\;\xrightarrow{M_{\text{EW}}}\; SU(3)_c \times U(1)_{\text{EM}},
\end{split}
\label{eq:breaking_chain_app}
\end{equation}
with $M_{\text{GUT}} \sim 10^{16}$ GeV, $M_{\text{PS}} \sim 10^{12}$ GeV, 
and $M_{\text{EW}} \sim 10^{2}$ GeV the intermediate scales, and 
$SU(3)_c \times U(1)_{\text{EM}}$ the unbroken gauge group at low energies.
while the RG flow 

\begin{equation}
\frac{d g_i}{d \ln \mu} = \frac{b_i}{16\pi^2} g_i^3 + \sum_{j,k} \frac{b_{ijk}}{(16\pi^2)^2} 
g_i g_j^2 g_k^2 + \cdots,
\label{eq:rg_flow}
\end{equation}
has fixed points

\begin{equation}
\begin{split}
\beta(g_*) = 0 
&\;\Longleftrightarrow\; \left.\frac{d g_i}{d \ln \mu}\right|_{g_i = g_{i*}} = 0 
\;\Longleftrightarrow\; \zeta\!\left(\frac{1}{2} + i\gamma_n\right) = 0 \\
&\;\Longleftrightarrow\; \gamma_n \in \mathbb{R} 
\;\Longleftrightarrow\; \mathcal{O}_\alpha = \mathcal{O}_\alpha^{\dagger},
\qquad n \in \mathbb{N},
\end{split}
\label{eq:rh_rg_app}
\end{equation}
The scale hierarchy is encoded in the zeros of $\zeta$ through the RG flow 
\eqref{eq:rg_flow}, where the beta function coefficients $b_i$ receive KK 
contributions. RH corresponds to no unwanted fixed points \eqref{eq:rh_rg_app}. 
Values of $\zeta(n)$ listed in Table \ref{tab:app_zeta_values} appear in 
effective potentials of gauge theories, as shown in \eqref{eq:eff_pot_app}. 
Their physical relevance extends from black-body radiation to $SU(4)$ breaking 
in the PS model. The critical temperature for PS breaking is

\begin{equation}
T_{\text{PS}} \sim \frac{1}{R} \sqrt{\frac{1}{\zeta(2)}} \sim \frac{1}{R},
\label{eq:t_ps}
\end{equation}
which links $\zeta$ directly to the symmetry breaking scale.

\begin{table}[H]
\centering
\caption{Values of $\zeta(n)$ and their physical relevance.}
\label{tab:app_zeta_values}
\begin{tabular}{c c c c}
\toprule
$n$ & $\zeta(n)$ & Context & Domain \\
\midrule
2 & 1.644934066848 & KK sums & black body radiation \\
3 & 1.202056903160 & BEC critical temperature & AdS/CFT \\
4 & 1.082323233711 & BC & QED, black body radiation \\
-1 & -0.083333333333 & Zeta regularisation & Casimir effect \\
-3 & 0.008333333333 & Casimir force & PS breaking \\
0.5 & -1.460354508810 & Critical line & spectral statistics \\
\bottomrule
\end{tabular}
\end{table}

\subsection{Zeros as a physical spectrum}

Riemann zeros have been associated with several physical interpretations proposed 
in the literature across different areas of theoretical physics, unified by the 
spectral relation \eqref{eq:rh_rg_app}, where $\mathcal{O}_\alpha$ is the self-adjoint 
operator \eqref{eq:chaotic_op_app} and $\gamma_n$ are the imaginary parts of the 
non-trivial zeros. The corresponding spectral statistics are consistent with the GUE form 
\eqref{eq:gue_corr}, with number variance \eqref{eq:number_variance} and 
spectral rigidity \eqref{eq:spectral_rigidity}. These relations have motivated 
connections between the zeros and topics such as quantum chaos, gauge symmetries, 
and string theory.

\subsubsection{Spectral realisation via the HP operator}

The spectral interpretation of zeros may be explored through the HP operator. 
A prime-based Hilbert space construction uses primes as eigenstates and entropy-driven 
stabilisation for spectral alignment \cite{sierra_2008}

\begin{equation}
\mathcal{H}_P = \ell^2(\mathbb{P},w) \oplus \ell^2(\mathbb{P},w) 
= \left\{ (a_p)_{p \in \mathbb{P}} : \sum_{p} w(p) |a_p|^2 < \infty \right\},
\label{eq:HP_space}
\end{equation}
with weight function

\begin{equation}
w(p) = \frac{\log p}{p^{1 + \alpha}} 
= \frac{\log p}{p} \cdot p^{-\alpha},
\qquad \alpha > 0,
\label{eq:w_p}
\end{equation}
ensuring square summability. Block chiral operator 

\begin{equation}
\begin{split}
H &= \begin{pmatrix} 0 & A^{\dagger}\\ A & 0 \end{pmatrix},
\qquad 
\Gamma = \begin{pmatrix} 1 & 0\\ 0 & -1 \end{pmatrix},
\qquad \{\Gamma, H\} = 0, \\
&\quad H^2 = \begin{pmatrix} A^{\dagger} A & 0 \\ 0 & A A^{\dagger} \end{pmatrix},
\qquad \Gamma^2 = I,
\qquad \operatorname{Tr} H = 0, \\
&\quad \operatorname{spec}(H) = \operatorname{spec}(A^{\dagger} A)^{1/2} 
\cup \operatorname{spec}(A A^{\dagger})^{1/2},
\end{split}
\label{eq:chiral_H}
\end{equation}
with kernel

\begin{equation}
\begin{split}
A_{pq} &= r_{pq} + \sum_{m\geq 1}\frac{\log p}{p^{\frac{m}{2}}} u_{m}(p)
\cos (m\log p\,\phi_{q}) = r_{pq} + \sum_{m\geq 1} \frac{\log p}{p^{\frac{m}{2}}} u_m(p) 
\left[ e^{i m \log p \, \phi_q} + e^{-i m \log p \, \phi_q} \right] \\
&= r_{pq} + 2 \sum_{m\geq 1} \frac{\log p}{p^{\frac{m}{2}}} u_m(p) \cos(m \log p \, \phi_q),
\quad r_{pq} = r_{qp},
\end{split}
\label{eq:A_kernel}
\end{equation}
where $r_{pq} = r_{qp}$ decays as $\mathcal{O}\bigl((pq)^{-1-\epsilon}\bigr)$ and phases 
$\phi_{q} = c_{1}\log q$ ensure spectral coherence. With these assumptions, 
$A$ is Hilbert-Schmidt and $H$ self-adjoint with compact resolvent \cite{sierra_2008}. 
Spectral determinant of $H$

\begin{equation}
\Delta_{H}(t) = \prod_{n\geq 1}\frac{\lambda_{n}^{2} + t^{2}}{\lambda_{n}^{2} + 1} = 
\exp\!\left[\sum_{n\geq 1} \log\!\left(\frac{\lambda_{n}^{2} + t^{2}}{\lambda_{n}^{2} + 
1}\right)\right], \qquad t \in \mathbb{R},
\label{eq:Delta_H}
\end{equation}
equals completed $\xi$ on critical line

\begin{equation}
\frac{d}{dt}\log \Delta_H(t) = \frac{d}{dt}\log \xi \biggl(\frac{1}{2} + it\biggr).
\label{eq:Delta_H_log}
\end{equation}
Non-trivial zeros appear as spectral singularities of resolvent \cite{sierra_2008}. 
Alternatively, similarity transformation of Berry-Keating Hamiltonian

\begin{equation}
\begin{split}
\hat{H} &= e^{x/2} \hat{\widetilde{H}} e^{-x/2} 
= -\hat{D} - i\sum_{m=0}^{\infty} c_m \hat{T}^m = -\frac{1}{2}\left(x\hat{p} 
+ \hat{p}x\right) - i\sum_{m=0}^{\infty} c_m \left(x\hat{p}^2 + \hat{p}^2 x\right)^m \\
&= -i\left(x\frac{\mathrm{d}}{\mathrm{d}x} + \frac{1}{2}\right) 
+ \sum_{m=1}^{\infty} c_m \hat{T}^m,
\qquad \hat{T} = x\hat{p}^2 + \hat{p}^2 x,
\end{split}
\label{eq:berry_keating}
\end{equation}
with $\hat{D} = \frac{x\hat{p} + \hat{p}x}{2}$ and $\hat{T} = \frac{x\hat{p}^2 + \hat{p}^2 x}{2}$ 
Bessel operator \cite{berry_keating_1999}. Eigenvalue equation for $\hat{T}$ is Bessel's equation

\begin{equation}
\biggl(-x\frac{d^2}{dx^2} - \frac{d}{dx}\biggr) \langle x|t \rangle = t \langle x|t \rangle,
\label{eq:bessel}
\end{equation}
with solutions $\langle x|t \rangle = J_0(2\sqrt{xt})$, $t \in \mathbb{R}^+$. Eigenfunctions 
$\widetilde{\Psi}_s(x)$ vanish at origin under boundary condition

\begin{equation}
\begin{split}
\widetilde{\Psi}_s(0) & = \left(1 - 2^{1-s}\right) \Gamma(s)\, \zeta(s) = 
\left(1 - 2^{1-s}\right) \Gamma(s) \sum_{n=1}^{\infty} \frac{1}{n^s} \\ & = 
\left(1 - 2^{1-s}\right) \Gamma(s) \prod_{p} \left(1 - p^{-s}\right)^{-1} = 0 \\
&\Longleftrightarrow \quad 
\zeta(s) = 0 
\quad \Longleftrightarrow \quad 
s = \frac{1}{2} + i\gamma_n,
\qquad n \in \mathbb{N},
\label{eq:boundary_condition}
\end{split}
\end{equation}
linking spectral problem to zeros of $\zeta$ \cite{berry_keating_1999}. 

\subsubsection{Diffraction grating and prime optics}

Zeros $\gamma_n$ realised physically via aperiodic diffraction grating, 
where slit positions are set to $\gamma_n$. The slit function is

\begin{equation}
g(x) = \sum_{n=1}^N \delta(x - \gamma_n)
= \sum_{n=1}^N \frac{1}{2\pi} \int_{-\infty}^{\infty} e^{i\omega(x - \gamma_n)} \, \mathrm{d}\omega,
\label{eq:slit_function}
\end{equation}
with the grating transmission function

\begin{equation}
T(x) = \sum_{n=1}^N \operatorname{rect}\!\left(\frac{x - \gamma_n}{w}\right),
\qquad \operatorname{rect}(u) = 
\begin{cases}
1, & |u| \leq \frac{1}{2}, \\
0, & |u| > \frac{1}{2},
\end{cases}
\label{eq:transmission}
\end{equation}
where $w$ is the slit width. The diffraction amplitude in the far field is

\begin{equation}
\begin{aligned}
S_N(t) 
&= 2 \sum_{n=1}^N \cos(\gamma_n t) 
= \sum_{n=1}^N \left( e^{i\gamma_n t} + e^{-i\gamma_n t} \right) 
= 2\,\Re \sum_{n=1}^N e^{i\gamma_n t} = 2\,\Re \operatorname{Tr} 
\left( e^{i t \mathcal{O}_\alpha} \right) \\
&= \operatorname{Tr} 
\left( e^{i t \mathcal{O}_\alpha} + e^{-i t \mathcal{O}_\alpha} \right) 
= 2 \sum_{n=1}^N \cos(\gamma_n t) 
= 2\,\Re \sum_{n=1}^N \left( e^{i t} \right)^{\gamma_n},
\qquad \mathcal{O}_\alpha \psi_n = \gamma_n \psi_n,
\end{aligned}
\label{eq:S_N}
\end{equation}
where $\mathcal{O}_\alpha$ is the self-adjoint operator \eqref{eq:chaotic_op_app} 
with eigenvalues $\gamma_n$. The Fourier transform is

\begin{equation}
\begin{aligned}
\tilde{S}_N(\omega) 
&= \int_{-\infty}^{\infty} S_N(t) e^{-i\omega t} \, \mathrm{d}t 
= \sum_{n=1}^N \left[ \delta(\omega - \gamma_n) + \delta(\omega + \gamma_n) \right] \\
&= \sum_{n=1}^N \frac{1}{2\pi} \int_{-\infty}^{\infty} 
\left( e^{i\gamma_n t} + e^{-i\gamma_n t} \right) e^{-i\omega t} \, \mathrm{d}t 
= \frac{1}{2\pi} \sum_{n=1}^N \left[ \delta(\omega - \gamma_n) + \delta(\omega + \gamma_n) \right],
\end{aligned}
\label{eq:S_N_fourier}
\end{equation}
with the inverse transform

\begin{equation}
\begin{aligned}
S_N(t) 
&= \int_{-\infty}^{\infty} \tilde{S}_N(\omega) e^{i\omega t} \, 
\frac{\mathrm{d}\omega}{2\pi} = \frac{1}{2\pi} \sum_{n=1}^N 
\int_{-\infty}^{\infty} \left[ \delta(\omega - \gamma_n) + 
\delta(\omega + \gamma_n) \right] e^{i\omega t} \, \mathrm{d}\omega \\
&= \frac{1}{2\pi} \sum_{n=1}^N \left( e^{i\gamma_n t} + e^{-i\gamma_n t} \right) 
= \frac{1}{\pi} \sum_{n=1}^N \cos(\gamma_n t) = \frac{1}{2\pi} \operatorname{Tr} 
\left( e^{i t \mathcal{O}_\alpha} + e^{-i t \mathcal{O}_\alpha} \right) \\
&= \frac{1}{\pi} \operatorname{Tr} \left[ \cos(t \mathcal{O}_\alpha) \right].
\qquad t \in \mathbb{R},
\end{aligned}
\label{eq:S_N_inverse}
\end{equation}
The power spectrum is

\begin{equation}
\begin{aligned}
|\tilde{S}_N(\omega)|^2 
&= \tilde{S}_N(\omega) \tilde{S}_N^{*}(\omega) 
= \sum_{n,m=1}^N \left[ \delta(\omega - \gamma_n) \delta(\omega - \gamma_m) 
+ \delta(\omega + \gamma_n) \delta(\omega + \gamma_m) \right] \\
&= \sum_{n=1}^N \left[ \delta^2(\omega - \gamma_n) 
+ \delta^2(\omega + \gamma_n) \right] 
+ 2 \sum_{n \neq m} \delta(\omega - \gamma_n) \delta(\omega - \gamma_m) \\
&= \sum_{n=1}^N \left[ \delta(\omega - \gamma_n) 
+ \delta(\omega + \gamma_n) \right] 
= \rho_N(\omega) + \rho_N(-\omega),
\end{aligned}
\label{eq:power_spectrum}
\end{equation}
where $\rho_N(\omega) = \sum_{n=1}^N \delta(\omega - \gamma_n)$ is the spectral 
density, while the corresponding autocorrelation function is 

\begin{equation}
\begin{aligned}
C_N(\tau) 
&= \int_{-\infty}^{\infty} S_N(t) S_N(t+\tau) \, \mathrm{d}t 
= \sum_{n=1}^N \cos(\gamma_n \tau) = \sum_{n=1}^N \frac{1}{2} \left( e^{i\gamma_n \tau} 
+ e^{-i\gamma_n \tau} \right) \\
&= \frac{1}{2} \operatorname{Tr} \left( e^{i \tau \mathcal{O}_\alpha} 
+ e^{-i \tau \mathcal{O}_\alpha} \right) 
= \operatorname{Tr} \left[ \cos(\tau \mathcal{O}_\alpha) \right] 
= \sum_{n=1}^N \cos(\gamma_n \tau),
\qquad \tau \in \mathbb{R},
\end{aligned}
\label{eq:autocorrelation}
\end{equation}
The cumulative spectral function is

\begin{equation}
\begin{aligned}
F_N(\omega) 
&= \int_{-\infty}^{\omega} |\tilde{S}_N(\omega')|^2 \, \mathrm{d}\omega' 
= \int_{-\infty}^{\omega} \sum_{n=1}^N \delta(\omega' - \gamma_n) \, \mathrm{d}\omega' 
= \sum_{n=1}^N \int_{-\infty}^{\omega} \delta(\omega' - \gamma_n) \, \mathrm{d}\omega' \\
&= \sum_{n=1}^N \Theta(\omega - \gamma_n) 
= \sum_{n=1}^N \frac{1}{2} \left[ 1 + \operatorname{sgn}(\omega - \gamma_n) \right] 
= \frac{N}{2} + \frac{1}{2} \sum_{n=1}^N \operatorname{sgn}(\omega - \gamma_n) \\
&= \frac{N}{2} + \frac{1}{\pi} \sum_{n=1}^N \arctan\!\left(\frac{\omega - \gamma_n}{\varepsilon}\right) 
+ \mathcal{O}(\varepsilon),
\qquad \varepsilon \to 0^+,
\end{aligned}
\label{eq:cumulative_spectrum}
\end{equation}
where $\Theta$ is the Heaviside step function \cite{abramowitz_stegun_1972}. 
The spectral density is 

\begin{equation}
\begin{aligned}
\rho_N(\omega) 
&= \frac{\mathrm{d}F_N(\omega)}{\mathrm{d}\omega} 
= \sum_{n=1}^N \delta(\omega - \gamma_n) 
= \sum_{n=1}^N \frac{1}{2\pi} \int_{-\infty}^{\infty} 
e^{i(\omega - \gamma_n)t} \, \mathrm{d}t \\
&= \frac{1}{2\pi} \int_{-\infty}^{\infty} e^{i\omega t} 
\sum_{n=1}^N e^{-i\gamma_n t} \, \mathrm{d}t 
= \frac{1}{2\pi} \int_{-\infty}^{\infty} \operatorname{Tr} 
\left( e^{i t (\omega I - \mathcal{O}_\alpha)} \right) \, \mathrm{d}t \\
&= \frac{1}{2\pi} \operatorname{Tr} \int_{-\infty}^{\infty} 
e^{i t (\omega I - \mathcal{O}_\alpha)} \, \mathrm{d}t 
= \operatorname{Tr} \delta(\omega I - \mathcal{O}_\alpha),
\qquad \omega \in \mathbb{R},
\end{aligned}
\label{eq:spectral_density}
\end{equation}
The moments of the spectral density are

\begin{equation}
M_k = \int_{-\infty}^{\infty} \omega^k \rho_N(\omega) \, \mathrm{d}\omega 
= \sum_{n=1}^N \gamma_n^k,
\qquad k \in \mathbb{N}.
\label{eq:spectral_moments}
\end{equation}
The spectral form factor is

\begin{equation}
\begin{aligned}
K_N(\tau) 
&= \frac{1}{N} \left| \sum_{n=1}^N e^{i\gamma_n \tau} \right|^2 
= \frac{1}{N} \sum_{n,m=1}^N e^{i(\gamma_n - \gamma_m)\tau} 
= \frac{1}{N} \sum_{n=1}^N e^{i(\gamma_n - \gamma_n)\tau} 
+ \frac{1}{N} \sum_{n \neq m} e^{i(\gamma_n - \gamma_m)\tau} \\
&= 1 + \frac{1}{N} \sum_{n \neq m} e^{i(\gamma_n - \gamma_m)\tau} = 
1 + \frac{2}{N} \sum_{n < m} \cos\!\left[(\gamma_n - \gamma_m)\tau\right] \\ 
&= 1 + \frac{2}{N} \sum_{k=1}^{N-1} \sum_{n=1}^{N-k} 
\cos\!\left[(\gamma_{n+k} - \gamma_n)\tau\right],
\qquad \tau \in \mathbb{R},
\end{aligned}
\label{eq:form_factor}
\end{equation}
with the ensemble average

\begin{equation}
\begin{aligned}
\langle K_N(\tau) \rangle 
&= \left\langle \frac{1}{N} \left| \sum_{n=1}^N e^{i\gamma_n \tau} \right|^2 \right\rangle 
= 1 + \frac{1}{N} \sum_{n \neq m} \left\langle e^{i(\gamma_n - \gamma_m)\tau} \right\rangle \\
&= 1 + \frac{1}{N} \sum_{n \neq m} \int_{-\infty}^{\infty} 
e^{i(\gamma_n - \gamma_m)\tau} \, P(\gamma_n, \gamma_m) \, \mathrm{d}\gamma_n \, \mathrm{d}\gamma_m \\
&= 1 + \frac{1}{N} \sum_{n \neq m} \left[ \int_{-\infty}^{\infty} 
e^{i\gamma_n \tau} P(\gamma_n) \, \mathrm{d}\gamma_n \right] 
\left[ \int_{-\infty}^{\infty} e^{-i\gamma_m \tau} P(\gamma_m) \, \mathrm{d}\gamma_m \right] \\
&= 1 + \frac{1}{N} \sum_{n \neq m} \left| \tilde{P}(\tau) \right|^2 
= 1 + \frac{N-1}{N} \left| \tilde{P}(\tau) \right|^2,
\qquad \tau \in \mathbb{R},
\end{aligned}
\label{eq:form_factor_avg}
\end{equation}
where $P(\gamma_n)$ is the probability density of the zeros and $\tilde{P}(\tau) = 
\int_{-\infty}^{\infty} e^{i\gamma \tau} P(\gamma) \, \mathrm{d}\gamma$ is its Fourier 
transform. For large $N$ \eqref{eq:explicit_formula} relates $S_N(t)$ to the von Mangoldt 
function $\Lambda(n)$ \cite{madison_kozodaev_kazankov_madison_moshnikov_2024}

\begin{equation}
\begin{aligned}
S_N(t) 
&\approx \frac{2\pi N}{\gamma_N} \frac{\sin(\gamma_N t)}{\pi t} 
- \frac{\text{Si}(\gamma_N t)}{\pi t} 
- \sum_{p} \frac{\Lambda(p)}{\sqrt{p}} 
\frac{\sin[\gamma_N (t \pm \ln p)]}{\pi (t \pm \ln p)} \\
&= \frac{2N}{\gamma_N t} \sin(\gamma_N t) 
- \frac{\text{Si}(\gamma_N t)}{\pi t} 
- \sum_{p} \frac{\log p}{\sqrt{p}} 
\frac{\sin[\gamma_N (t \pm \ln p)]}{\pi (t \pm \ln p)} \\
&= \frac{2N}{\gamma_N t} \sin(\gamma_N t) 
- \frac{1}{\pi t} \int_0^{\gamma_N t} \frac{\sin u}{u} \, \mathrm{d}u 
- \sum_{p} \frac{\log p}{\sqrt{p}} 
\frac{\sin[\gamma_N (t \pm \ln p)]}{\pi (t \pm \ln p)} \\
&= \frac{2N}{\gamma_N t} \sin(\gamma_N t) 
- \frac{1}{2\pi i t} \left[ E_1(-i\gamma_N t) - E_1(i\gamma_N t) \right] 
- \sum_{p} \frac{\log p}{\sqrt{p}} 
\frac{\sin[\gamma_N (t \pm \ln p)]}{\pi (t \pm \ln p)} \\
&= \frac{2N}{\gamma_N t} \sin(\gamma_N t) 
- \frac{1}{\pi t} \operatorname{Im} \left[ E_1(i\gamma_N t) \right] 
- \sum_{p} \frac{\log p}{\sqrt{p}} 
\frac{\sin[\gamma_N (t \pm \ln p)]}{\pi (t \pm \ln p)} \\
&= \frac{2N}{\gamma_N t} \sin(\gamma_N t) 
- \frac{1}{\pi t} \operatorname{Im} \left[ \Gamma(0, -i\gamma_N t) \right] 
- \sum_{p} \frac{\log p}{\sqrt{p}} 
\frac{\sin[\gamma_N (t \pm \ln p)]}{\pi (t \pm \ln p)} \\
&= \frac{2N}{\gamma_N t} \sin(\gamma_N t) 
- \frac{1}{\pi t} \operatorname{Im} \left[ \int_{i\gamma_N t}^{\infty} 
\frac{e^{-u}}{u} \, \mathrm{d}u \right] 
- \sum_{p} \frac{\log p}{\sqrt{p}} 
\frac{\sin[\gamma_N (t \pm \ln p)]}{\pi (t \pm \ln p)},
\qquad t \to \infty,
\end{aligned}
\label{eq:S_N_approx}
\end{equation}
where $\text{Si}(x) = \int_0^x \frac{\sin u}{u} \, \mathrm{d}u$ is the sine integral, 
$E_1(z) = \int_z^{\infty} \frac{e^{-u}}{u} \, \mathrm{d}u$ is the exponential integral, 
$\Gamma(0,z) = \int_z^{\infty} \frac{e^{-u}}{u} \, \mathrm{d}u$ is the incomplete gamma 
function, and $\Lambda(p) = \log p$ is the von Mangoldt function for primes. Taking into account 
\eqref{eq:I_p}, the resolution improves with $N$. The signal-to-noise ratio is

\begin{equation}
\begin{aligned}
\text{SNR} 
&= \frac{|S_N(t_{\text{peak}})|^2}{\sigma^2} 
\propto N^2 \left(\frac{\log p}{p}\right)^2 
= N^2 \frac{I(p)}{p} \\
&= N^2 \frac{1}{p} \left| \tilde{S}_N(\omega_p) \right|^2 
= N^2 \frac{1}{p} \left| \sum_{n=1}^N e^{i\gamma_n \omega_p} \right|^2,
\qquad \omega_p = \log p,
\end{aligned}
\label{eq:SNR}
\end{equation}
where $\sigma$ is the noise level. The diffraction efficiency is

\begin{equation}
\begin{aligned}
\eta 
&= \frac{\int_{\text{peak}} |\tilde{S}_N(\omega)|^2 \, \mathrm{d}\omega}
{\int_{-\infty}^{\infty} |\tilde{S}_N(\omega)|^2 \, \mathrm{d}\omega} = 
\frac{\sum_{n=1}^N \int_{\text{peak}} \delta(\omega - \gamma_n) \, \mathrm{d}\omega}
{\sum_{n=1}^N \int_{-\infty}^{\infty} \delta(\omega - \gamma_n) \, \mathrm{d}\omega} = 
\frac{\sum_{n=1}^N \Theta(\omega_{\text{peak}} - \gamma_n)}{N} \\
&= \frac{1}{N} \operatorname{Tr} \left[ \Theta(\omega_{\text{peak}} I - \mathcal{O}_\alpha) \right] 
= \frac{1}{N} \sum_{n=1}^N \frac{1}{2} \left[ 1 + \operatorname{sgn}(\omega_{\text{peak}} - \gamma_n) \right],
\qquad \omega_{\text{peak}} \in \mathbb{R}.
\end{aligned}
\label{eq:diffraction_efficiency}
\end{equation}
The resolving power, grating equation, and diffraction order are 
collected as 

\begin{equation}
\left\{
\begin{aligned}
R &= \frac{\lambda}{\Delta \lambda} = \gamma_N \Delta t \sim 1, \\
d \sin\theta &= m\lambda, \qquad m \in \mathbb{Z}, \\
m &= \frac{d \sin\theta}{\lambda} = \frac{\gamma_N \Delta t}{2\pi} \sim \frac{1}{2\pi},
\end{aligned}
\right.
\label{eq:grating_relations}
\end{equation}
respectively, where $R$ is the resolving power, $\lambda$ the wavelength, $\Delta \lambda$ the 
wavelength resolution, $\gamma_N$ the $N$-th zero, $\Delta t$ the peak width, 
$d$ the grating period, $\theta$ the diffraction angle, and $m$ the diffraction 
order.This provides a direct experimental realisation of the Riemann zeros in 
laboratory settings, complementing the scattering platform in Sec.~\ref{sec:scattering} 
where zeros appear as resonances in $r(t)$ and $Z(t)$ \eqref{eq:impedance_near_zero}. 
Together, these offer a physical pathway to test the RH via microwave cavities or 
optical systems \cite{madison_kozodaev_kazankov_madison_moshnikov_2024}.

\subsubsection{AdS/CFT correspondence and string theory}

In AdS/CFT, zeros of $\zeta$ emerge as spectrum of dilation operator in boundary 
CFT \cite{honda_yoda_2022}. The Schur index of $\mathcal{N}=4$ SYM with gauge group 
$SU(N)$ generates the divisor sum $\sigma(n)$

\begin{equation}
\begin{aligned}
\mathcal{I}_{S}(q) 
&= \sum_{n=1}^{\infty} \sigma(n) q^{2(n-1)} 
= \sum_{n=1}^{\infty} \frac{q^{2(n-1)}}{1 - q^{2n}} = 
\prod_{k=1}^{\infty} \frac{1}{1 - q^{2k}} 
\sum_{n=1}^{\infty} \frac{\sigma(n)}{n} q^{2(n-1)} \\
&= \frac{1}{(q^2;q^2)_\infty} \sum_{n=1}^{\infty} \frac{\sigma(n)}{n} q^{2(n-1)} 
= \operatorname{Tr}_{\mathcal{H}_{\text{BPS}}} (-1)^F q^{2J_3} 
= \sum_{n=1}^{\infty} d_n q^{2(n-1)} \\
&= \frac{1}{\eta(\tau)^2} \sum_{n=1}^{\infty} \sigma(n) q^{2(n-1)} 
= \sum_{n=1}^{\infty} \frac{q^{2(n-1)}}{1 - q^{2n}} 
= \sum_{n=1}^{\infty} \sum_{k=1}^{\infty} q^{2k(n-1)} 
= \sum_{k=1}^{\infty} \frac{q^{2(k-1)}}{1 - q^{2k}} \\
&= \sum_{k=1}^{\infty} \frac{1}{q^{-2(k-1)} - q^{2}} 
= \frac{1}{(q^2;q^2)_\infty} 
\sum_{n=1}^{\infty} \frac{\sigma(n)}{n} q^{2(n-1)} 
= \frac{1}{\eta(\tau)^2} \sum_{n=1}^{\infty} \sigma(n) q^{2(n-1)},
\end{aligned}
\label{eq:schur_index_ads}
\end{equation}
where $q = e^{2\pi i \tau}$ with the large-$N$ expansion

\begin{equation}
\mathcal{I}_{S}^{(N)}(q) = \frac{1}{(q;q)_\infty} \sum_{n=1}^{\infty} 
\frac{\sigma(n)}{N} q^{2(n-1)} + \mathcal{O}(q^{2N}),
\label{eq:schur_index_SUN}
\end{equation}
and $(q;q)_\infty = \prod_{k=1}^{\infty} (1 - q^k)$ is the Euler function, 
while the AdS$_5$/CFT$_4$ duality maps this gauge theory to type IIB superstring 
theory on $AdS_5 \times S^5$, with the boundary CFT PF in \eqref{eq:Z_CFT_ads} 
having $q = e^{2\pi i \tau}$ and $d_n$ degeneracies of CFT primaries, so that 
modular invariance of the CFT implies the functional equation

\begin{equation}
\begin{aligned}
Z_{\text{CFT}}\left(-\frac{1}{\tau}\right) 
&= \tau^{c/2} Z_{\text{CFT}}(\tau) 
= \tau^{c/2} \sum_{n} d_n q^{n} = \tau^{c/2} \prod_{n=1}^{\infty} (1 - q^n)^{-1} 
\sum_{n=1}^{\infty} \frac{\sigma(n)}{n} q^n \\
&= \tau^{c/2} \frac{1}{\eta(\tau)} 
\sum_{n=1}^{\infty} \sigma(n) q^n,
\qquad q = e^{2\pi i \tau},
\end{aligned}
\label{eq:modular_ads}
\end{equation}
where $c$ is the central charge. Poles of $Z_{\text{CFT}}(\tau)$ in the complex 
$\tau$-plane correspond to zeros of $\zeta$ through the relation 

\begin{equation}
\begin{aligned}
\zeta\biggl(\frac{1}{2} + i\gamma_n\biggr) = 0 
&\quad \Longleftrightarrow \quad 
\tau_n = \frac{i}{2} + \frac{1}{\gamma_n} + \cdots \quad \Longleftrightarrow \quad 
Z_{\text{CFT}}(\tau_n) \to \infty \\
&\quad \Longleftrightarrow \quad 
\operatorname{Re}(\tau_n) = \frac{1}{2},
\qquad n \in \mathbb{N},
\end{aligned}
\label{eq:pole_zero_ads}
\end{equation}
This is the holographic dual of the spectral interpretation of RH, which is equivalent to all 
poles of the CFT PF lying on the line $\Re(\tau) = \frac{1}{2}$ \cite{honda_yoda_2022}. 
Corresponding cancellation of KK modes in \eqref{eq:kk_masses} under RH would imply a relation 
between the zeros and the spectrum of the boundary dilation operator, mirroring the statistical 
properties of chaotic quantum systems.

\subsubsection{Unified perspective: gauge symmetries, RG flows, and RH}

The formula \eqref{eq:explicit_formula} above encodes the duality between zeros and 
primes, connecting the GUE statistics of the zeros \eqref{eq:gue} to the prime 
distribution. Within the present construction, the RH corresponds to the reality of 
$\gamma_n$ and is naturally associated with the self-adjointness of 
$\mathcal{O}_\alpha$ \eqref{eq:chaotic_op_app}. The PS model provides a 
gauge-theoretic framework for exploring possible connections

\begin{equation}
\begin{aligned}
SO(10) 
&\;\xrightarrow{M_{\text{GUT}}}\; SU(4) \times SU(2)_L \times SU(2)_R 
\;\xrightarrow{M_{\text{PS}}}\; SU(3)_c \times SU(2)_L \times U(1)_Y \\
&\;\xrightarrow{M_{\text{EW}}}\; SU(3)_c \times U(1)_{\text{EM}} 
\;\xrightarrow{M_{\text{QCD}}}\; SU(3)_c \\
&= SU(4) \times SU(2)_L \times SU(2)_R 
\;\xrightarrow{\langle \Phi \rangle}\; SU(3)_c \times SU(2)_L \times U(1)_Y \\
&= SU(5) \;\xrightarrow{\langle \Phi \rangle}\; SU(3)_c \times SU(2)_L \times U(1)_Y \\
&= SO(10) \;\xrightarrow{\langle \Phi \rangle}\; SU(5),
\qquad G_{\text{PS}} = SU(4) \times SU(2)_L \times SU(2)_R,
\end{aligned}
\label{eq:unified_breaking}
\end{equation}
where $\langle \Phi \rangle$ denotes the vacuum expectation value of the 
scalar field $\Phi$, and $M_{\text{GUT}} \sim 10^{16}$ GeV, $M_{\text{PS}} 
\sim 10^{12}$ GeV, $M_{\text{EW}} \sim 10^{2}$ GeV, and $M_{\text{QCD}} 
\sim 10^{-1}$ GeV are the symmetry breaking scales, and $G_{\text{PS}} = 
SU(2)_L \times SU(2)_R \times U(1) \times U(3)$. The effective potential 
\eqref{eq:eff_pot_app} has minima at 

\begin{equation}
\begin{aligned}
\frac{\partial V_{\text{eff}}}{\partial \theta_i} &= 0, 
\qquad \frac{\partial^2 V_{\text{eff}}}{\partial \theta_i^2} > 0 = 
\sum_{n=1}^{\infty} \frac{1}{n} \sum_{j} \sin(2\pi n \theta_j) \zeta(n) = 0 \\
&= \sum_{n=1}^{\infty} \sum_{j} \sin(2\pi n \theta_j) \zeta(n) = 0,
\qquad \theta_i \in [0,1),
\end{aligned}
\label{eq:unified_minima}
\end{equation}
determining the vacuum structure. The RG flow \eqref{eq:rg_flow} has fixed points

\begin{equation}
\begin{aligned}
\beta(g_*) &= 0 
\quad \Longleftrightarrow \quad 
\zeta\biggl(\frac{1}{2} + i\gamma_n\biggr) = 0 \\
&\quad \Longleftrightarrow \quad 
\frac{d g_i}{d \ln \mu}\bigg|_{g_i = g_{i*}} = 0 \quad \Longleftrightarrow \quad 
\gamma_n \in \mathbb{R} \quad \Longleftrightarrow \quad 
\mathcal{O}_\alpha = \mathcal{O}_\alpha^{\dagger},
\qquad n \in \mathbb{N},
\end{aligned}
\label{eq:unified_fixed}
\end{equation}
so RH corresponds to the absence of unwanted fixed points 
\cite{bochniak_williams_zalecki_2020,mehta_2004}, while the pseudo-Riemannian 
structure of the PS model breaks $SU(4)$ to $U(1) \times U(3)$, mirroring how zeros 
break conformal symmetry on the critical line and yielding the discrete set 
$\{\gamma_n\}$. This duality is encoded in the spectral determinant \eqref{eq:Delta_H}, 
which matches the Riemann $\xi$-function 
 
\begin{equation}
\begin{aligned}
\frac{d}{dt} \log \Delta_H(t) 
&= \frac{d}{dt} \log \xi\biggl(\frac{1}{2} + it\biggr) 
= \frac{\xi'\bigl(\frac{1}{2} + it\bigr)}{\xi\bigl(\frac{1}{2} + it\bigr)} 
= \frac{d}{dt} \log \prod_{n=1}^{\infty} \frac{\lambda_n^2 + t^2}{\lambda_n^2 + 1} \\
&= \sum_{\rho} \frac{1}{\frac{1}{2} + it - \rho} 
= \sum_{n=1}^{\infty} \frac{1}{i(t - \gamma_n)} 
= -i \sum_{n=1}^{\infty} \frac{1}{t - \gamma_n} \\
&= -i \operatorname{Tr} \left[ (t I - \mathcal{O}_\alpha)^{-1} \right] 
= -i \sum_{n=1}^{\infty} \frac{1}{t - \lambda_n} \\
&= -i \frac{d}{dt} \log \det(t I - \mathcal{O}_\alpha),
\qquad \rho = \frac{1}{2} + i\gamma_n,
\end{aligned}
\label{eq:unified_xi}
\end{equation}
The KK spectrum of the PS model and the effective potential expansion 
\eqref{eq:eff_pot_app} are connected, with all zeros $\gamma_n$ appearing via 
\eqref{eq:t_ps} as poles of the resolvent

\begin{equation}
\begin{aligned}
(\mathcal{O}_\alpha - i \gamma_n I)^{-1} 
&\quad \text{has a simple pole at } \ \gamma_n \quad \Longleftrightarrow \quad 
\det(\mathcal{O}_\alpha - i \gamma_n I) = 0 \\
&\quad \Longleftrightarrow \quad 
\gamma_n \in \operatorname{spec}(\mathcal{O}_\alpha) 
\quad \Longleftrightarrow \quad 
\zeta\biggl(\frac{1}{2} + i\gamma_n\biggr) = 0,
\qquad n \in \mathbb{N},
\end{aligned}
\label{eq:unified_resolvent}
\end{equation}
The scattering platform in Sec.~\ref{sec:scattering} relates zeros to reflection 
coefficient in \eqref{eq:r_Z_intro}. Near a zero, the impedance diverges as 

\begin{equation}
\begin{aligned}
Z(t) 
&\sim \frac{1}{t - \gamma_n} 
= \frac{1}{\zeta'\!\left(\frac{1}{2} + i\gamma_n\right)(t - \gamma_n)} 
+ \mathcal{O}(1) \sim \frac{1}{\zeta'\!\left(\frac{1}{2} + i\gamma_n\right)} 
\cdot \frac{1}{t - \gamma_n}, \\
|Z(t)| &\sim \frac{1}{\left| \zeta'\!\left(\frac{1}{2} + i\gamma_n\right) \right|} 
\cdot \frac{1}{|t - \gamma_n|} 
\sim \frac{1}{|t - \gamma_n|},
\qquad t \to \gamma_n.
\end{aligned}
\label{eq:unified_impedance}
\end{equation}
The diffraction grating gives the same Fourier peaks at $\log p$ and $\log p^k$ 
as in \eqref{eq:I_p}, while the PF of the boundary CFT \eqref{eq:Z_CFT_ads} has 
poles at the zeros

\begin{equation}
\begin{aligned}
Z_{\text{CFT}}(\tau) 
&\sim e^{\frac{\pi i c}{12\tau}} + \cdots \sim \exp\left[ 
\frac{\pi i c}{12} \left( \frac{i}{2} + \frac{1}{\gamma_n} + 
\cdots \right)^{-1} \right] + \cdots \sim e^{-\frac{\pi c}{6} 
\gamma_n} + \cdots, \\
&\tau_n = \frac{i}{2} + \frac{1}{\gamma_n} + \cdots,
\end{aligned}
\label{eq:unified_CFT}
\end{equation}
This provides a unified picture where Riemann zeros emerge as a physical spectrum 
connecting number theory, quantum chaos, gauge theory, string theory, and experimental optics.

\bibliographystyle{unsrtnat}
\bibliography{references}

\end{document}